\documentclass[a4paper,11pt]{article}

\pdfoutput=1 % if your are submitting a pdflatex (i.e. if you have
\usepackage{jcappub} % for details on the use of the package, please
\newcommand{\Mpl}{{M_{p}}}

\title{\bf Observable Signatures of a Classical Transition}

\author[a,b]{Matthew C. Johnson,}
\author[a]{Wei Lin}

\affiliation[a]{Department of Physics and Astronomy, York University \\ Toronto, On, M3J 1P3, Canada}
\affiliation[b]{Perimeter Institute for Theoretical Physics \\ Waterloo, Ontario N2L 2Y5, Canada}

\emailAdd{mjohnson@perimeterinstitute.ca}

\abstract{Eternal inflation arising from a potential landscape predicts that our universe is one realization of many possible cosmological histories. One way to access different cosmological histories is via the nucleation of bubble universes from a metastable false vacuum. Another way to sample different cosmological histories is via classical transitions, the creation of pocket universes through the collision between bubbles. Using relativistic numerical simulations, we examine the possibility of observationally determining if our observable universe resulted from a classical transition. We find that classical transitions produce spatially infinite, approximately open Friedman-Robertson-Walker universes. The leading set of observables in the aftermath of a classical transition are negative spatial curvature and a contribution to the Cosmic Microwave Background temperature quadrupole. The level of curvature and magnitude of the quadrupole are dependent on the position of the observer, and we determine the possible range of observables for two classes of single-scalar field models. For the first class, where the inflationary phase has a lower energy than the vacuum preceding the classical transition, the magnitude of the observed quadrupole generally falls to zero with distance from the collision while the spatial curvature grows to a constant. For the second class, where the inflationary phase has a higher energy than the vacuum preceding the classical transition, the magnitude of the observed quadrupole generically falls to zero with distance from the collision while the spatial curvature grows without bound. We find that the magnitude of the quadrupole and curvature grow with increasing centre of mass energy of the collision, and explore variations of the parameters in the scalar field lagrangian.
}

\begin{document} 
\maketitle

%%%%%%%%%%%%%%%%%%%%%%%%%%%%%%%%%%%%%%%%%%%%%%
\section{Introduction} \label{secintro}
%%%%%%%%%%%%%%%%%%%%%%%%%%%%%%%%%%%%%%%%%%%%%%
In the picture of eternal inflation (see Refs.~\cite{Guth:2007ng,Guth:2013epa,Aguirre:2007gy} for a review), our observable universe is part of a multiverse in which different regions of space are occupied by different field values in a potential landscape (for example, the simple potential landscape shown in Fig.~ \ref{CTfigure}). In eternal inflation, a region of the universe initially occupying a false vacuum gives rise to bubble universes that ultimately reach different vacua after a period of cosmological evolution. If the rate of bubble formation does not outpace the background expansion rate, inflation becomes eternal and the process of bubble formation carries on forever.

Although the phase transition out of the false vacuum does not complete, all of the bubbles that form during eternal inflation undergo collisions with others~\cite{Guth:1982pn,Garriga:2006hw}. If our Observable Universe is contained within one such bubble, collisions provide a possible observational test of eternal inflation~\citep{Aguirre:2007an}. A substantial body of work has been devoted to determining the outcome of bubble collisions and predicting the observational signatures~\cite{Aguirre:2007an,Aguirre:2007wm,Aguirre:2008wy,Aguirre:2009ug,Feeney_etal:2010dd,Feeney_etal:2010jj,Feeney:2012hj,Johnson:2010bn,Johnson:2011wt,McEwen:2012uk,Wainwright:2013lea,Wainwright:2014pta,Zhang:2015uta,Chang_Kleban_Levi:2009,Chang:2007eq,Czech:2010rg,Freivogel_etal:2009it,Freivogel:2005vv,Gobbetti_Kleban:2012,Kleban_Levi_Sigurdson:2011,Kleban:2011pg,Salem:2012,Zhang:2015bga,Ahlqvist:2013whn,Amin:2013dqa,Amin:2013eqa,Easther:2009ft,Giblin:2010bd,Blanco-Pillado:2003hq,Deskins:2012tj,Freivogel:2007fx,Garriga:2006hw,Gott:1984ps,Hawking:1982ga,Hwang:2012pj,Kozaczuk:2012sx,Larjo:2009mt,Moss:1994pi,Osborne:2013hea,Osborne:2013jea,Wu:1984eda,Feeney:2015hwa}, and the first searches for the predicted pattern in the cosmic microwave background (CMB) anisotropy have put constraints on the number and amplitude of collision signatures in our observable Universe~\cite{Feeney_etal:2010jj,Feeney_etal:2010dd,Feeney:2012hj,Osborne:2013hea}. 

In a simple potential landscape with only two vacua, colliding bubbles merge. When the scalar field potential has three or more local minima, bubble collisions can produce domain walls separating regions occupying different vacua. In some cases, the collision can produce a region that occupies a new vacuum, distinct from the vacua in either of the colliding bubbles. Such an event is known as a classical transition~\cite{Easther:2009ft}, and is an alternative to quantum mechanical bubble nucleation as a mechanism for populating vacua and cosmological histories in a potential landscape. In some cases, classical transitions may represent the only mechanism for accessing certain regions of a potential landscape~\cite{Johnson:2010bn}. If our Universe inhabits such a corner of the potential landscape, then it must have arisen from a classical transition. In this case, the rate at which classical transitions occur is irrelevant, since bubble collisions are guaranteed during eternal inflation. However, in the absence of a more complete theory, it is difficult to judge how generic classical transitions might be.

As we will see below, the Universe to the future of a classical transition is approximately Friedman-Robertson-Walker (FRW) with negative curvature. In analogy to models of open inflation~\cite{Bucher:1994gb}, one can invoke a period of slow-roll inflation after the classical transition occurs in order to dilute this spatial curvature and produce the appropriate density perturbations to seed large scale structure. In this way, one can create models where our Universe might exist to the future of a classical transition. The aim of this paper is to investigate the possibility that our Observable Universe could have been produced by a classical transition, and compute the observational signatures of this scenario.

Because classical transitions are an intrinsically non-linear phenomenon, it is necessary to employ numerical simulations. To determine the cosmology to the future of a classical transition, we must perform these simulations in full General Relativity. Fortunately, the collision spacetime possesses a hyperbolic symmetry which allows us to simulate the entire collision spacetime using a simulation in one space and one time dimension. This symmetry makes performing simulations computationally cheap, allowing us to simulate many different cases. Throughout the paper, we use the simulation framework developed in Refs.~\cite{Wainwright:2013lea,Wainwright:2014pta,wainwright_new}. This framework directly links the scalar field landscape underlying eternal inflation to the predictions for cosmological observables, allowing for observational constraints to be placed on the theory underlying eternal inflation.

We find that classical transitions produce spatially infinite approximately FRW universes. The leading cosmological observables are negative spatial curvature and an approximately quadratic comoving curvature perturbation with planar symmetry. Such a curvature perturbation maps to a temperature quadrupole in the Cosmic Microwave Background radiation. The prediction for curvature and the CMB quadrupole are correlated, and depend on the distance of the observer (in an appropriate set of coordinates) from the spatial location of the collision that caused the classical transition. In addition to observer position, the initial separations between the two colliding bubbles is a random variable. In addition to depending on the scalar field lagrangian, the observables also depend on the initial separation. Therefore, any specific scalar field lagrangian giving rise to classical transitions yields an ensemble of predictions for observation determined by the initial separation of the colliding bubbles and the position of an observer. We classify the range of predictions allowed in the context of two classes of models for the scalar field potential.  

The plan of the paper is as follows. In Sec.~\ref{sec:classtrans} we review the mechanism of classical transitions and the expected geometries allowed in GR. In Sec.~\ref{sec:framework} we describe the simulation framework and methods for extracting cosmological observables. In Sec.~\ref{secsimresult} we present the results from our simulations, and in Sec.~\ref{secconclude} we conclude

%%%%%%%%%%%%%%%%%%%%%%%%%%%%%%%%%%%%%%%%
%%%%%%
%%%%%%
%%%%%%%%%%%%%%%%%%%%%%%%%%%%%%%%%%%%%%%%
%%%%%%
%%%%%%
%%%%%%%%%%%%%%%%%%%%%%%%%%%%%%%%%%%%%%%%
%%%%%%
%%%%%%

\section{Classical Transitions}\label{sec:classtrans}

\begin{figure}
\begin{center}
	\includegraphics[width=\textwidth]{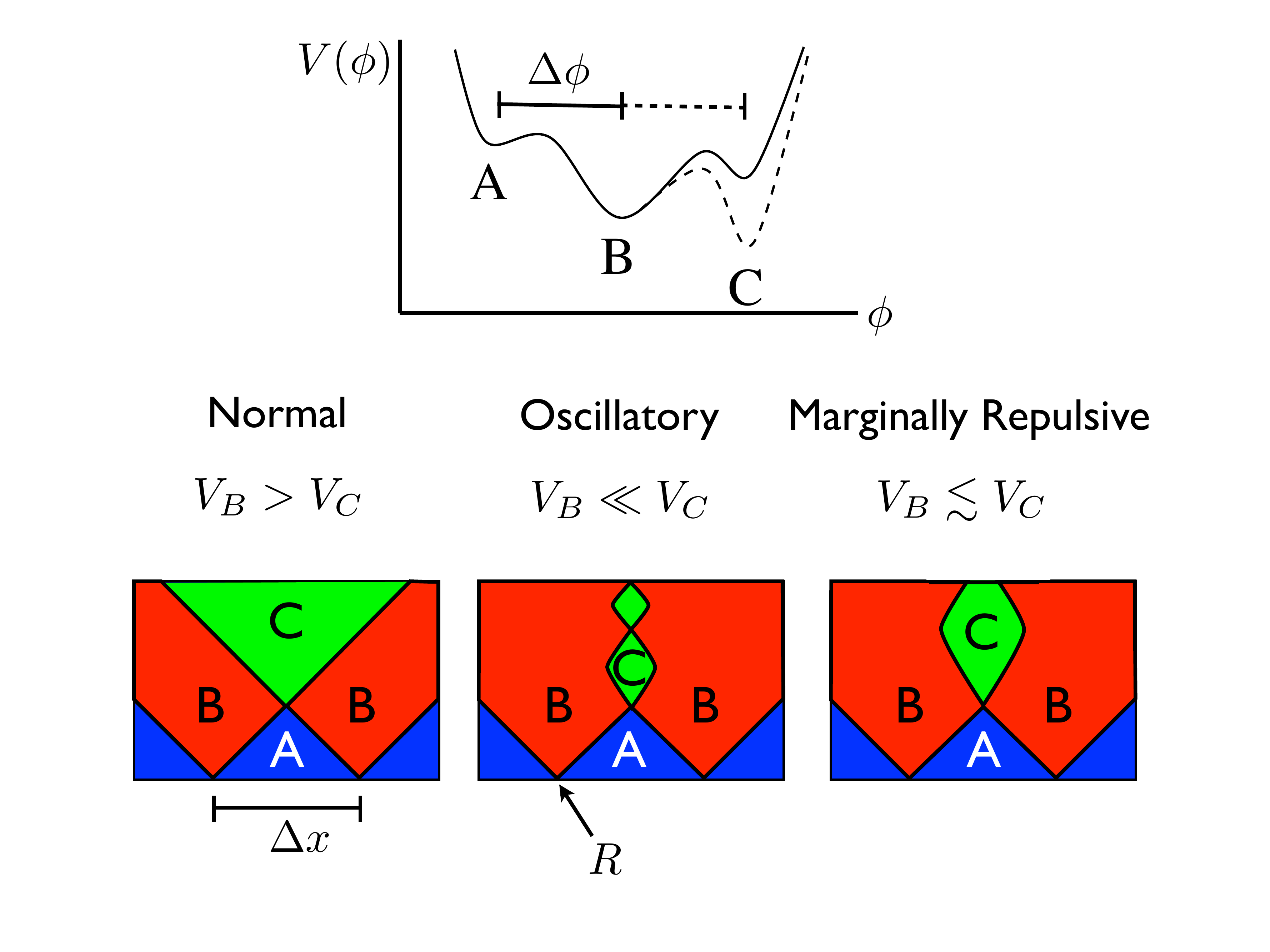}
	\caption{ A scalar field potential that gives rise to classical transitions (top) and the possible classical transition geometries (bottom). The scalar field potential has three local minima: $A$, $B$, and $C$. We will consider cases where the potential $V_C$ at $C$ is both lower (dashed) and higher (solid) than the potential $V_B$ at $B$. The field excursion between vacuum A and vacuum B is $\Delta \phi$. In the free passage approximation, the distance the field is pushed in the future of a collision between two bubbles of B embedded in a background of A is bounded by $\Delta \phi$. Including gravity, there are a number of distinct classical transition geometries (bottom). We consider bubbles of $B$ nucleated in a background false vacuum of $A$. Bubbles of $B$ form with an initial radius $R$, and in the centre of mass frame, are separated by a proper distance $\Delta x$. For a Normal geometry, which corresponds to $V_B > V_C$, the region of $C$ produced by the collision expands indefinitely. In an Oscillatory geometry, corresponding to $V_C \ll V_B$, the initially expanding region of $C$ eventually collapses, spawning a new region of $C$ which expands and contracts. Eventually this oscillation terminates in the release of scalar radiation, and vacuum $C$ is no longer present in the spacetime. Repulsive and Marginally Repulsive geometries occur when $V_B$ and $V_C$ are comparable. For $V_B \lesssim V_C$, a Marginally Repulsive geometry results, in which the region of $C$ expands but is prevented from contracting by the gravitational field of the domain wall and the positive energy of vacuum $C$. This gives rise to a lasting region of $C$. }\label{CTfigure}
	\end{center}
\end{figure}

In Fig.~\ref{CTfigure}, we outline a number of properties of classical transitions. Consider a potential with local minima A, B, and C. We will consider cases where the potential $V_C$ at $C$ is both lower (dashed) and higher (solid) than the potential $V_B$ at $B$. Vacuum A is a false vacuum and bubbles of B spontaneously form through quantum tunnelling, mediated by the Coleman and De Luccia (CDL)~\citep{cdl,fate1,fate2}. Many collisions between bubbles of $B$ occur, in spite of the fact that the phase transition does not complete in eternal inflation. For pairwise collisions, which are most relevant in the regime of small nucleation rates, we are able to work in the centre of mass frame, where the two colliding bubbles nucleate at the same coordinate time~\cite{Aguirre:2007wm}. In the absence of gravity, the lorentz factor of the bubble walls at the time of collision is given by:
\begin{equation}\label{eq:lorentzfactor}
\gamma = \frac{\Delta x}{R},
\end{equation}
where $R$ is the initial radius of the colliding bubbles and $\Delta x$ is the proper distance between the bubbles in the centre of mass frame.  The lorentz factor characterizes the centre of mass energy of the collision. 

When the bubbles of $B$ collide, the field in the immediate aftermath of the collision is displaced. If the centre of mass energy is large enough ($\gamma \sim \mathcal{O}(1)$), this displacement reaches a maximum size, given by the superposition of two bubble profiles~\cite{Giblin:2010bd}. In the limit of infinite boost, superposition holds exactly; this is known as the free passage approximation. For vacuum bubbles, as we consider here, the amplitude of each of the bubble profiles, and therefore the maximum field excursion, is the width of the potential barrier, $\Delta \phi$, separating $A$ and $B$. When the structure of the potential is such that the field is pushed into a local minimum, as in Fig.~\ref{CTfigure}, a classical transition results, producing a spacetime region containing phase $C$.

The ultimate fate of the spacetime region containing $C$ depends on the vacuum energies of the three phases. This is depicted in Fig.~\ref{CTfigure}. For $V_B > V_C$, the region of $C$ expands due to the outward pressure gradient across the domain wall separating $C$ from $B$. The result is a lasting region of $C$; this is known as a Normal geometry. Without gravity, when $V_B \leq V_C$, an Oscillatory geometry is produced. In an Oscillatory geometry, an initially expanding pocket of $C$ re-collapses due to the inward pressure gradient across the domain wall separating $C$ from $B$. The collision makes a new region of $C$ which itself re-collapses, and this process repeats until the colliding walls dissipate into scalar radiation. No lasting region of $C$ is produced in an Oscillatory geometry. Including gravitational effects~\cite{escapecrunch,Johnson:2011wt}, it is possible to produce a lasting region of $C$ even when $V_B \leq V_C$. These geometries are referred to as Marginally Repulsive, and occur when the gravitationally repulsive effects of the domain walls separating $C$ from $B$ are strong enough to prevent the region of $C$ from re-collapsing. This can occur when $V_B \lesssim V_C$, with $V_C$ bounded by the constraint~\footnote{Note that Eq.~\ref{eq:mrceq} implies that $V_C < V_A$ for a marginally repulsive geometry. Therefore, one can never attain a potential higher than the False vacuum by a classical transition.}
\begin{equation}\label{eq:mrceq}
V_C \leq \frac{(V_A+V_B)^2}{4V_A}.
\end{equation}
In this paper, we focus on Normal and Marginally Repulsive geometries, both of which contain lasting regions of $C$.

Previous work has characterized and simulated classical transitions~\cite{Easther:2009ft,Giblin:2010bd,Johnson:2010bn,Johnson:2011wt,Amin:2013dqa,Amin:2013eqa}. The focus of the present work is on determining what the cosmological observables would be if we inhabited a region produced by a classical transition, a question to which we now turn.

\section{Simulation Framework}\label{sec:framework}
\subsection{Scalar field potential}\label{secmodel}
We use a quadratic inflationary potential with two Gaussian bumps for our model potential landscape:
\begin{equation}\label{V_(phi)}
V(\phi)=A_1Exp\left[-\frac{\phi^2}{2 \Delta\phi_1^2}\right]\pm A_2Exp\left[-\frac{(\phi-\sigma)^2}{2 \Delta\phi_2^2}\right]+\frac{1}{2}m^2(\phi-\phi_0)^2.
\end{equation}
In Eq. \ref{V_(phi)}, $``+"/``-"$ indicates whether the second barrier is above (AIP) or below (BIP) the inflationary plateau respectively. The inflationary component of the potential is necessary to dilute the curvature of spatial slices in the classical transition spacetime, as we will see below. From the discussion in the previous section, the AIP models produce classical transitions with Normal geometries and the BIP models produce Marginally Repulsive geometries. This choice of potential has no microphysical motivation, but has enough freedom to explore the phenomenology of classical transitions. $\Delta\phi_{1/2}$ are the widths of the bumps, $\sigma$ is the shift of the second bump from the first one, and $\phi_0$ specifies the location of the true vacuum. Following Ref.~\citep{Wainwright:2014pta}, we choose to parameterize $A_1$ and $A_2$ by:
\begin{equation}
A_1\equiv\beta_1|m^2\phi_0\Delta\phi_1e^{1/2}|, \ \ \ A_2\equiv\beta_2|m^2\phi_0\Delta\phi_2e^{1/2}|
\end{equation}
In general, there are five free variables associated with the potential Eq.~\ref{V_(phi)}: the barrier widths $(\Delta\phi_{1/2})$, the parametrized bump heights $(\beta_{1/2})$ and the shift $(\sigma)$. Below, we will measure $\Delta\phi_{1/2}$ and $\sigma$ in terms of $\, \Mpl$; $\beta_{1/2}$ is dimensionless. 

\begin{figure}
	\centering
	\includegraphics[width=.6\textwidth]{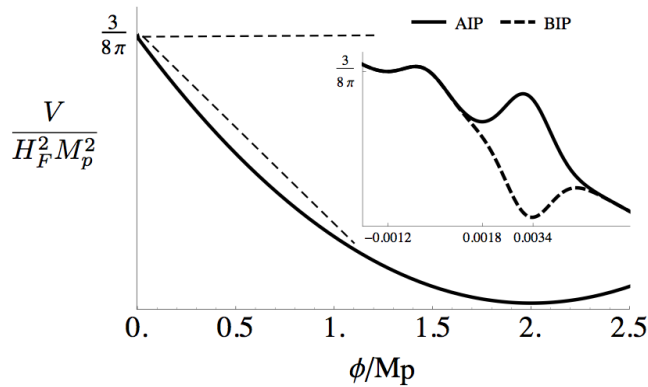}
	\caption{The potential Eq.~\ref{V_(phi)} for a characteristic choice of parameters. The true vaccum is located at $\phi=2 \, \Mpl$, the solid line is the AIP model; the dashed line is the BIP model. }\label{CTpotential}
\end{figure}

We first discuss the parameters of the AIP model. We can make a few simplifications to reduce the dimensionality of parameter space. The free passage approximation suggests that the structure of the second barrier (characterized by $\Delta\phi_{2}$ and $\beta_2$) is largely irrelevant in the AIP model. Free passage also suggests that there will be large parameter degeneracies, as the most important determinants of the outcome of a collision are the initial profile of the bubble wall (given by the analytic continuation of the CDL instanton) and the initial separation between bubbles. 

\begin{figure}
\begin{center}
         \includegraphics[width=.32\textwidth]{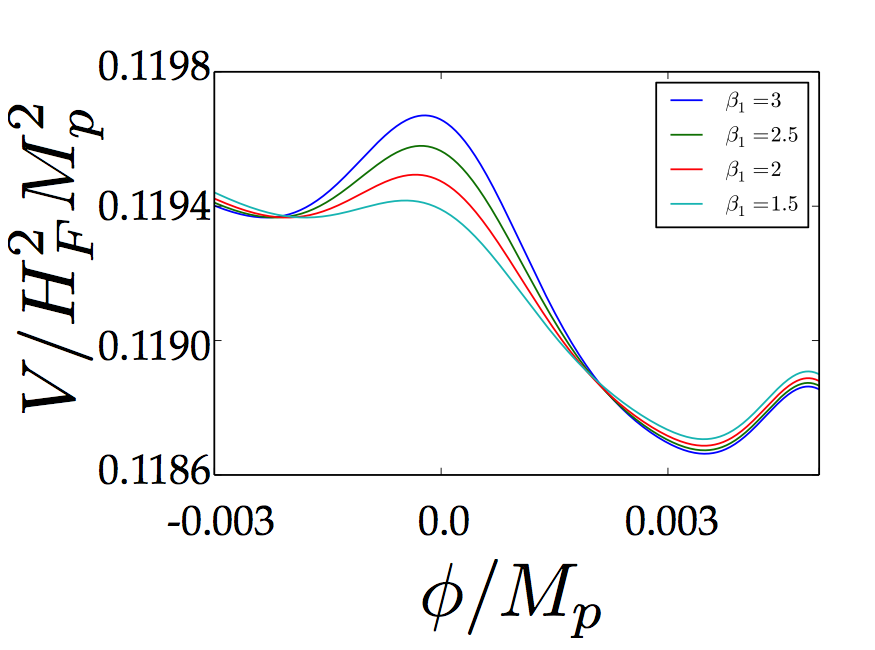}
	\includegraphics[width=.32\textwidth]{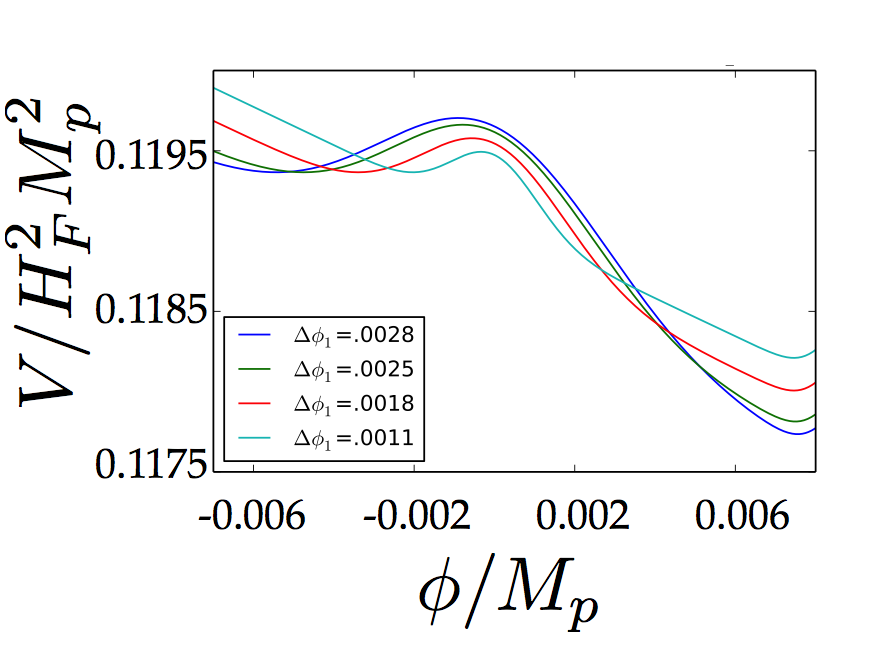}
        	\includegraphics[width=.32\textwidth]{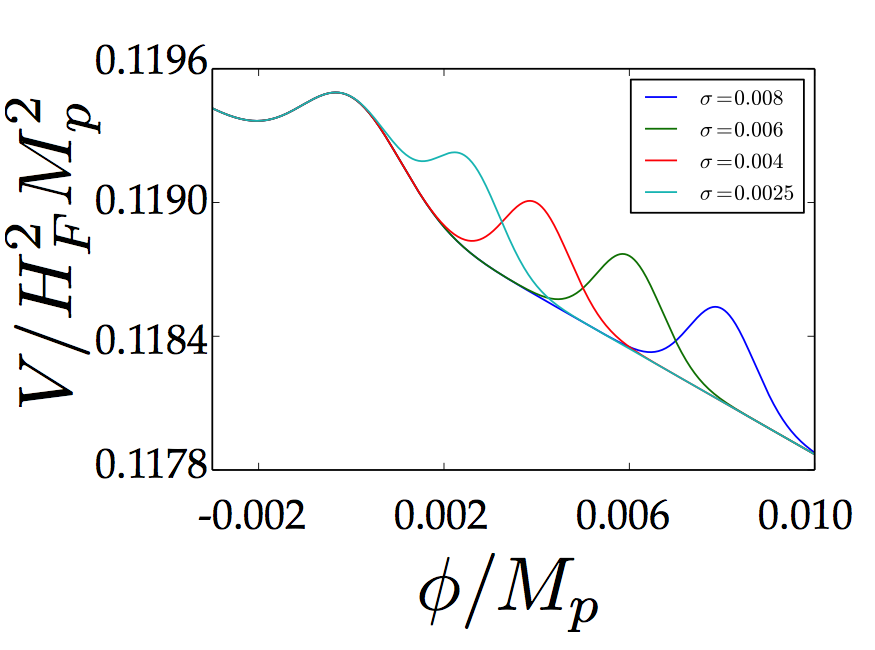}
        \caption{Potential Eq.~\ref{CTpotential} with varying parameters $\beta_{1}$ (left panel), $\Delta\phi_{1}$ (centre panel), and $\sigma$ (right panel).}\label{vhws}
\end{center}
\end{figure}

\begin{figure}
\begin{center}
        		\includegraphics[width=.32\textwidth]{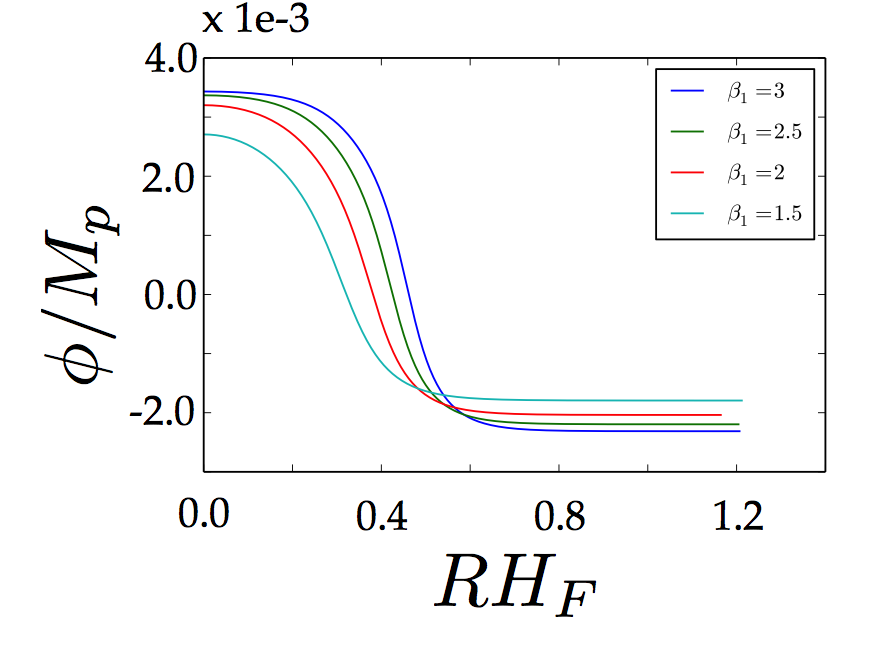}
		\includegraphics[width=.32\textwidth]{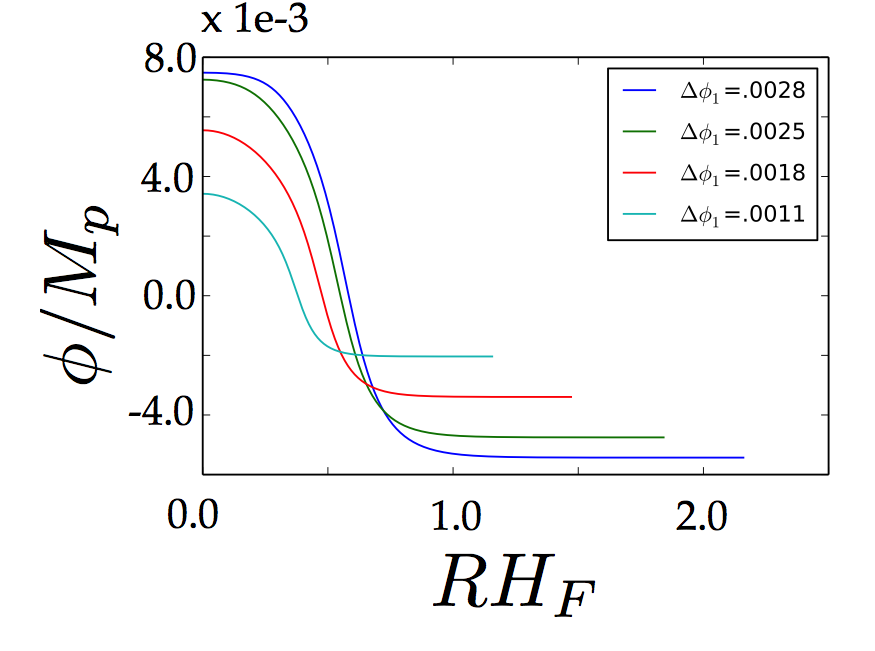}
		\includegraphics[width=.32\textwidth]{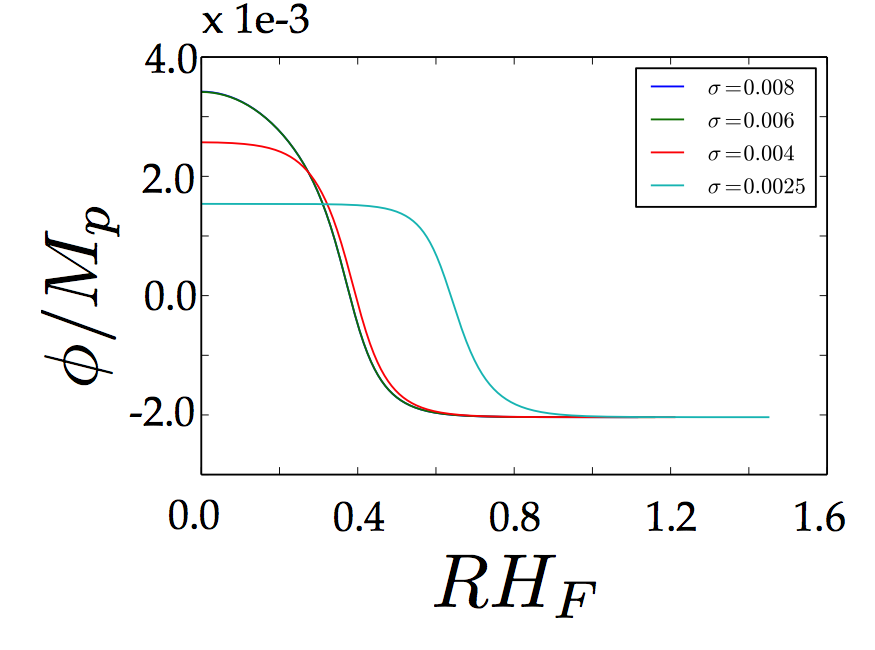}
        \caption{Instanton profiles for the potentials shown in Fig \ref{vhws}: varying $\beta_{1}$ (left panel), $\Delta\phi_{1}$ (centre panel), and $\sigma$ (right panel).}\label{iphws}
\end{center}
\end{figure}

In Fig.~\ref{vhws}, we show the potential for the AIP model for varying $\beta_{1}$, $\Delta\phi_{1}$, and $\sigma$; the associated instanton profiles are shown in Fig.~\ref{iphws}. We hold $\beta_{2}$ and $\Delta\phi_{2}$ fixed for the AIP model. However, note that the first barrier can affect the shape of the second barrier in some regions of parameter space. 

In light of the free passage approximation, we can predict that the degeneracies in parameter space are associated with parameter combinations that give nearly identical instanton profiles. As a crude model, we can approximate instanton profiles by their amplitude $\phi_\mathrm{amp}$, defined as the distance in field space between the instanton endpoints, and their size $R$, defined as the position where the field attains the midpoint between the instanton endpoints. In the analysis below, we choose combinations of the parameters $\beta_{1}$, $\Delta\phi_{1}$, and $\sigma$ that give instanton profiles with a constant radius but varying amplitude, and vice versa. We can pair $\Delta\phi_1$, $\beta_1$ and $\sigma$ in three different combinations ($\Delta\phi_1-\beta_1$, $\Delta\phi_1-\sigma$ and $\beta_1-\sigma$) to vary the radius and amplitude. However, the window we are able to simulate in the $\Delta\phi_1-\beta_1$ sector is too narrow to show significant differences in $R$, and so we omit this sector in the analysis below.

\begin{figure}
\begin{center}
\includegraphics[width=.5\textwidth]{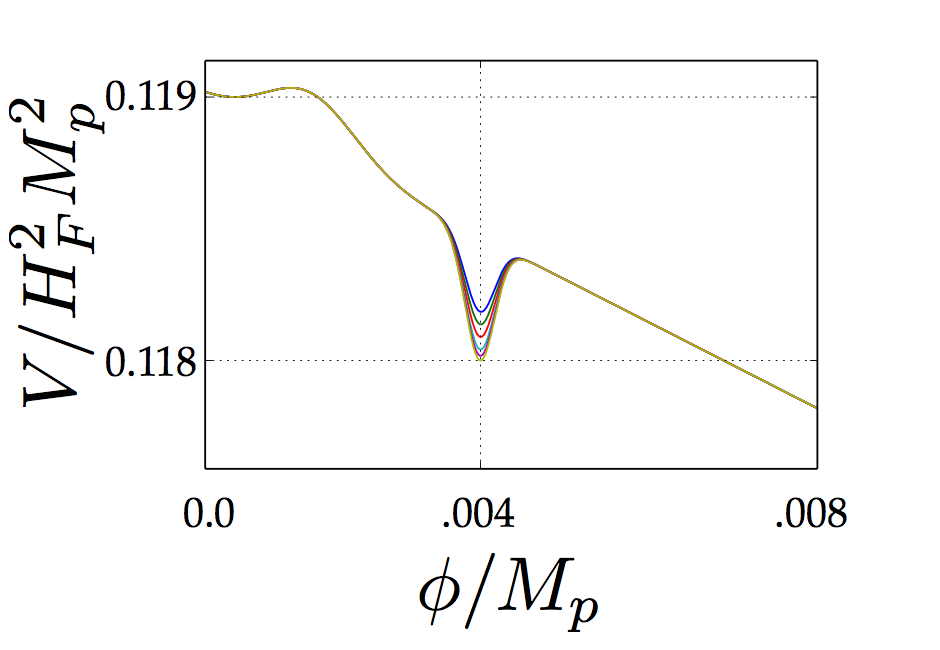}
\caption{$V(\phi)$ for the BIP model at increasing values of $\beta_2$, from $\beta_2 = 6$ (blue) to $\beta_2 = 9.87$ (yellow). The depth of the intermediate potential minimum increases with increasing $\beta_2$.}\label{m6p}
\end{center}
\end{figure}

For the BIP model, we are primarily interested in exploring the observables in Marginally Repulsive geometries. To this end, from Eq.~\ref{eq:mrceq}, we wish to change the relative vacuum energies in such a way to produce a Marginally Repulsive geometry. Varying only $\beta_2$, as shown in Fig.~\ref{m6p}, there will be a very small range of values for $\beta_2$ that produce a marginally repulsive geometry (recall that we must satisfy the inequality Eq.~\ref{eq:mrceq}). Simultaneously varying $\sigma-\beta_2$ or $\Delta\phi_2-\beta_2$ can produce a variety of cases in which a marginally repulsive geometry is produced, since varying $\sigma$ and/or $\Delta \phi_2$ will shift the energy density at the onset of slow-roll.

%%%%%%%%%%%%%%%%%%%%%%%%%%%%%%%%%%%%%%%%
%%%%%%
%%%%%%
%%%%%%%%%%%%%%%%%%%%%%%%%%%%%%%%%%%%%%%%
%%%%%%
%%%%%%
%%%%%%%%%%%%%%%%%%%%%%%%%%%%%%%%%%%%%%%%
%%%%%%
%%%%%%

\subsection{Collision Spacetime}
Individual CDL bubbles possess an $SO(3,1)$ symmetry~\cite{fate2,fate1,cdl}. The interior of each bubble is described by an open FRW metric, and its wall expands asymptotically to the speed of light. We will consider pair-wise collisions between bubbles; in the eternally inflating regime, this is sufficient to describe the region of spacetime causally accessible to most observers. The spacetime describing the collision between two CDL bubbles has an $SO(2,1)$ symmetry, and the metric can, in full generality, be written as~\citep{Wainwright:2013lea}:
\begin{equation}\label{eq:collisionmetric}
H_F^2ds^2 = -\alpha^2(N,x) dN^2+a^2(N,x) \cosh^2 N dx^2 + \sinh^2 N (d\chi^2+\sinh^2\chi \ d\psi^2)
\end{equation}
where $H_F$ is the false vacuum Hubble parameter, $N$ measures the numbers of $e$-foldings in the false vacuum, $x$ is the direction between the two bubble centers, and ${\chi, \psi}$ parameterize the hyperbolic symmetry of the collision spacetime.

Due to the hyperbolic symmetry, the collision spacetime can be obtained by doing simulations in one time and one space direction. The initial conditions are specified by the analytic continuation of the CDL instanton, and the unknown metric functions $\alpha(N,x)$ and $a(N,x)$ as well as the scalar field $\phi(N,x)$ evolved using the Einstein and scalar field equations. Importantly, the only free parameter in the initial conditions is the initial separation between the bubbles in the centre of mass frame. A complete description of the procedure for determining the initial value problem and subsequent evolution is described in Refs.~\cite{Wainwright:2013lea,Wainwright:2014pta,Johnson:2011wt}.

\subsection{Extracting CMB Observables from the Simulation}
The metric Eq.~\ref{eq:collisionmetric} and field $\phi(N,x)$ fully specify the collision spacetime. To make connection with cosmological observables, it is necessary to go from this global description to local descriptions of the Universe that would be seen by an ensemble of observers. There are a variety of approaches to this problem~\cite{Wainwright:2013lea,Wainwright:2014pta,wainwright_new,Xue:2013bva}. We use the method outlined in Ref.~\cite{wainwright_new}, which we briefly summarize here. 
\begin{itemize}
\item In the comoving gauge, the scalar field is homogeneous. Therefore, we can extract spatial slices in the comoving gauge by finding the surfaces of constant field. We choose a reference field value of $\phi_0 =0.15\, \Mpl$, which is sufficiently far into inflation that the comoving curvature perturbation is frozen in on large scales.
\item To coordinatize the surface of constant field, we use the proper distance $u$ from a fiducial point along the surface of constant field, and retain the hyperbolic symmetry of the collision spacetime. Projecting the metric Eq.~\ref{eq:collisionmetric} yields the spatial metric:
\begin{equation}
\label{eq:metric2}
H_F^2 ds_3^2 = du^2 + \sinh^2 N(u) (d\chi^2 + \sinh^2\chi d\varphi^2).
\end{equation}
\item We then define a reference point $x_0$, and make the linear coordinate transformation
\begin{equation}
u-u_0 = a_0 (\xi - \xi_0)
\end{equation}
which takes us to the perturbed hyperbolic slicing of FRW:
\begin{equation}
\label{eq:metric3}
H_F^2 ds^2 = a_0^2 [d\xi^2 + (1-2B)\cosh^2 \xi (d\rho^2 + \sinh^2\rho d\varphi^2)]
\end{equation}
We can identify $\xi_0$ as the coordiante associated with $x_0$ and $a_0$ as the scale factor at the reference point. 
\item We determine the scalar curvature of the metric Eq.~\ref{eq:metric3}, and integrate the definition of the comoving curvature perturbation:
\begin{equation}\label{eq:R}
a_0^2 R^{(3)}(\xi) = -6 + 4 \nabla^2\mathcal{R} (\xi | x_0).
\end{equation}
with the boundary conditions
\begin{equation}
\mathcal{R} (\xi_0|x_0) = 0, \ \ \ \partial_\xi \mathcal{R} (\xi_0|x_0) = 0.
\end{equation}
Neither $\mathcal{R} (\xi_0|x_0)$ nor $\partial_\xi \mathcal{R} (\xi_0|x_0)$ are physical observables, and so we are free to set them to zero.
\item We then transform to the anisotropic hyperbolic slicing of FRW introduced in Ref.~\cite{Wainwright:2013lea}
\begin{equation}
\label{eq:Rmetric}
H_F^2 ds^2 = a_0^2 (1 - 2 \mathcal{R}) [d\tilde{\xi}^2 + \cosh^2 \tilde{\xi} (d\rho^2 + \sinh^2\rho d\varphi^2)]
\end{equation}
In the vicinity of the reference point $\xi_0$, we can make the approximation that $\xi \simeq \tilde{\xi}$; the leading correction is given by
\begin{equation}
\tilde{\xi} = \xi + \frac{\partial_\xi^2 \mathcal{R} (\xi_0|x_0)}{6} (\xi - \xi_0)^3 
\end{equation}
which will be small in the window corresponding to our Observable Universe where $\xi - \xi_0 \ll 1$. In the following, we will substitute $\tilde{\xi} \rightarrow \xi$.
\item In the anisotropic hyperbolic coordinates, we perform a simple translation taking the point $\xi_0$ to the origin of coordinates $\xi=0$. 
\item Finally, we can find the comoving curvature perturbation as a function of Cartesian coordinates
\begin{equation}
\label{eq:Rmetric}
H_F^2 ds^2 = a_0^2 \frac{(1 - 2 \mathcal{R})}{(1 - \frac{R^2}{4 R_{\rm curv}^2})^2} [dX^2 + dY^2 + dZ^2]
\end{equation}
where $R \equiv X^2 + Y^2 + Z^2$. In the limit of small $\xi$, along the $\xi$-direction we have
\begin{equation}
R^2 = R_{\rm curv}^2 (\xi^2 + \mathcal{O}(\xi^4))
\end{equation}
and
\begin{equation}
X \simeq R_{\rm curv} \xi \cos \theta
\end{equation}
We therefore identify $\xi = 1$ with the curvature radius $R_{\rm curv}$. The curvature radius can be written in terms of the present energy density in curvature, $\Omega_k$, as:
\begin{equation}
R_{\rm curv} = \frac{1}{H_0 \sqrt{\Omega_k}}
\end{equation} 
where $H_0$ is the present day Hubble constant. The current constraint on spatial curvature from the Planck satellite, $\Omega_k=0.000 \pm 0.005$~\citep{planck}, implies that the curvature radius is at least $15 H_0^{-1}$, and therefore that an observer will have causal access to regions of size $\xi-\xi_0 \ll 1$.
\item This procedure is then repeated for an ensemble of reference points $x_0$, thereby tiling any constant field hypersurface with an ensemble of locally perturbed FRW patches.
\end{itemize}

To compare our results with the CMB, we need to relate temperature anisotropies to the comoving curvature perturbation in the classical transition spacetime. From the vantage point of any observer, the perturbation associated with a classical transition covers the entire sky. In addition, since the perturbation is a pre-inflationary relic, we expect the affect to be primarily on the largest scales in the CMB. Different observers in the classical transition spacetime are sampled by considering different reference points $x_0$. As we will see in the next section, for classical transitions there is a one-to-one mapping between $x_0$ and $\xi_0$. Below, we will use $\xi_0$ to label reference points. 

Working in the Sachs Wolfe approximation, the temperature fluctuation and $\mathcal{R}$ are related by \citep{sw}
\begin{equation}\label{tempfluc}
\frac{\Delta T}{T}=\frac{\mathcal{R}(\vec{X}_{\rm ls}|x_0)}{5},
\end{equation}
where $\vec{X}_{\rm ls}$ is the intersection of the past light cone of an observer with the surface of last scattering in cartesian coordinates, a 2-sphere of radius
\begin{equation}
R_{\rm ls}=2 \sqrt{\Omega_k (\xi_0)} R_{\rm curv}
\end{equation}
where $\Omega_k$ is the energy density in curvature today seen by an observer at $\xi_0$. In the following, it will be convenient to compute the relative curvature in different patches. We therefore express $\Omega_k(\xi_0)$ relative to $\Omega_k(0) \equiv \Omega_k(\xi_0=0)$:
\begin{equation}\label{omegaxi0}
\Omega_k(\xi_0)=\Omega_k(0) \frac{a_0^2(0)}{a_0^2(\xi_0)},
\end{equation} 
where $\xi_0=0$ is an arbitrary choice. The value of $\Omega_k(0)$ is set by the details of the slow-roll inflation model to the future of the classical transition, and we will leave it arbitrary in what follows.

Assuming that it is smooth (confirmed below), we can Taylor expand $\mathcal{R}$ about $X=0$ (equivalently $\xi=\xi_0$) and substitute with $X = R_{\rm ls} \cos \theta$ to obtain  
\begin{equation}\label{tempflucseries}
\frac{\Delta T}{T}\approx\frac{1}{5}\left[2 \partial_\xi^2 \mathcal{R}(\xi_0|x_0) \Omega_k(\xi_0) \cos^2 \theta + \mathcal{O}(\Omega_k^{3/2}(\xi_0)) \right].
\end{equation}
Expanding in spherical harmonics
\begin{equation}\label{basis}
\frac{\Delta T}{T}=\sum_\ell \sum_m a_{\ell m}Y_{\ell m},
\end{equation}
we can identify
\begin{equation}
a_{20} = \frac{8}{15} \sqrt{\frac{\pi}{5}} \partial_\xi^2 \mathcal{R}(\xi_0|x_0) \Omega_k(\xi_0).
\end{equation}
Substituting with Eq.~\ref{omegaxi0}, the contribution to the quadrupole of the angular power spectrum~\footnote{We report $\mathcal{D}_2$, defined as $\mathcal{D}_\ell = \ell(\ell+1)C_\ell/2\pi$, where $\langle a_{\ell m} a_{\ell' m'}\rangle = \delta_{\ell \ell'} \delta_{m m'} C_{\ell}$.} is therefore given by
\begin{equation}\label{dl}
\frac{\mathcal{D}_2}{\Omega_k(0)^2} = \frac{64}{375} \left( \partial_\xi^2 \mathcal{R}(\xi_0|x_0)\right)^2 \frac{a_0^4(0)}{a_0^4(\xi_0)}
\end{equation}

In summary, the leading observables in a classical transition spacetime are a correlated contribution to the spatial curvature (given by Eq.~\ref{omegaxi0}) and the quadrupole of the CMB temperature (given by Eq.~\ref{dl}), both expressed with respect to a model-dependent  reference curvature $\Omega_k(0)$.

\section{Simulation Results}\label{secsimresult}

\begin{figure}
\begin{center}
                \includegraphics[height=2.75in]{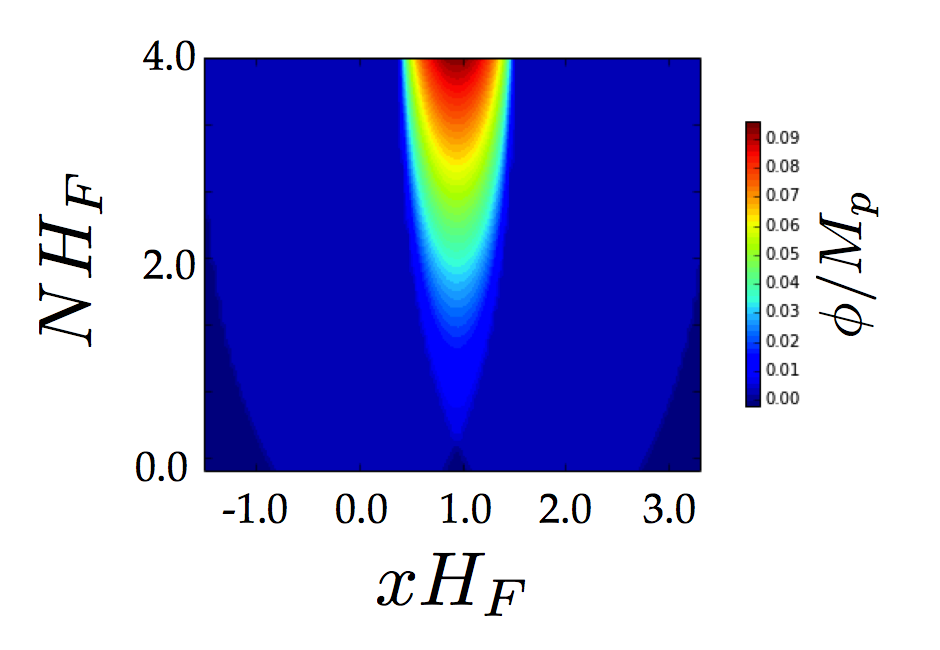}
        \caption{A contour plot of $\phi$ for a collision that produces a successful classical transition in the AIP model. The inflationary region of the potential is at $\phi > .005 \, \Mpl$.}\label{fig:AIPmodel}
\end{center}
\end{figure}

\begin{figure}
	\centering
	\includegraphics[width=.5\textwidth]{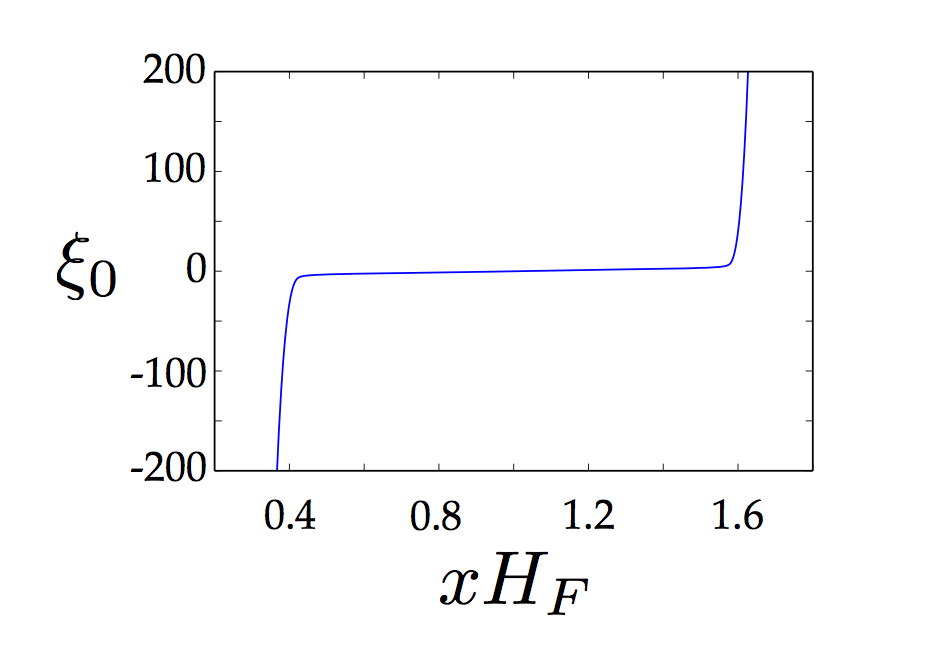}
	\caption{Anisotropic hyperbolic coordinate, $\xi$, as a function of spatial simulation coordinate, $H_F x$. $\xi$ is anti-symmetric about the location of the collision ($xH_F=1$), and diverges near the bubble wall. }\label{xix}
\end{figure}

Let us begin by applying the pipeline described above to one of the AIP models. In Fig.~\ref{fig:AIPmodel} we show a successful classical transition in an AIP model with the parameters $\Delta \phi_1 = 7.1 \times 10^{-4} \, \Mpl$, $\beta_1=1.5$, $\sigma=0.004\, \Mpl$. The two expanding bubbles collide near $x=1.0 H_F^{-1}$, creating a region of spacetime where the field rolls down the inflationary plateau. Choosing the constant field surface $\phi_0 = 0.15\, \Mpl$, we show the mapping between a reference simulation co-ordinate $x_0$ and the reference anisotropic hyperbolic FRW coordinate $\xi_0$ in Fig.~\ref{xix}. The coordinate map $\xi_0$ is one-to-one with $x_0$, anti-symmetric about the location of the collision, and diverges near the bubble wall. The wall reaches a constant comoving position in $x$, compressing the spatial extent of the slice in terms of $\xi_0$.  

\begin{figure}
\begin{center}
                \includegraphics[width=.49\textwidth]{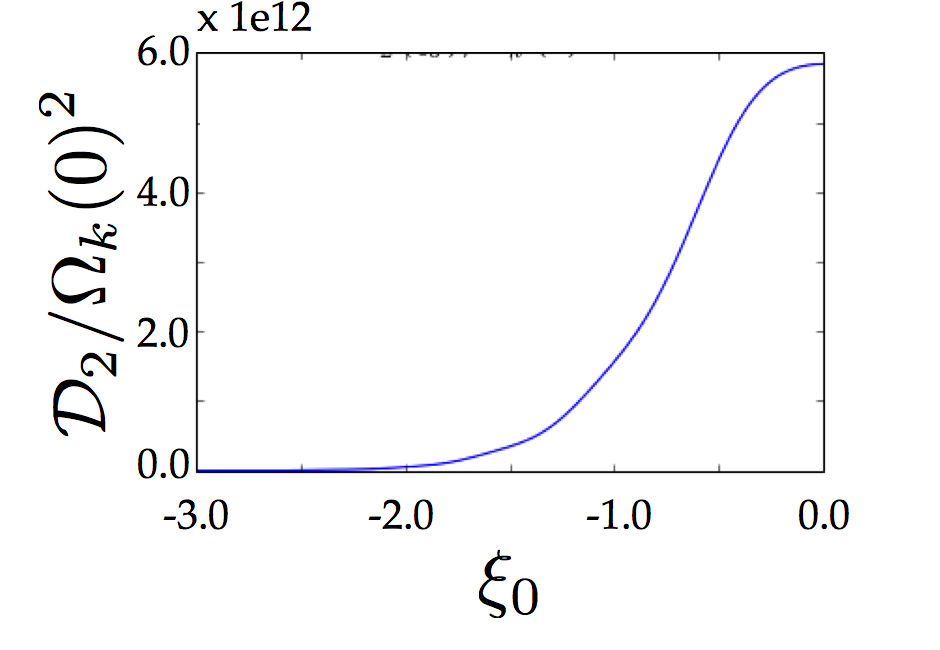}
                \includegraphics[width=.49\textwidth]{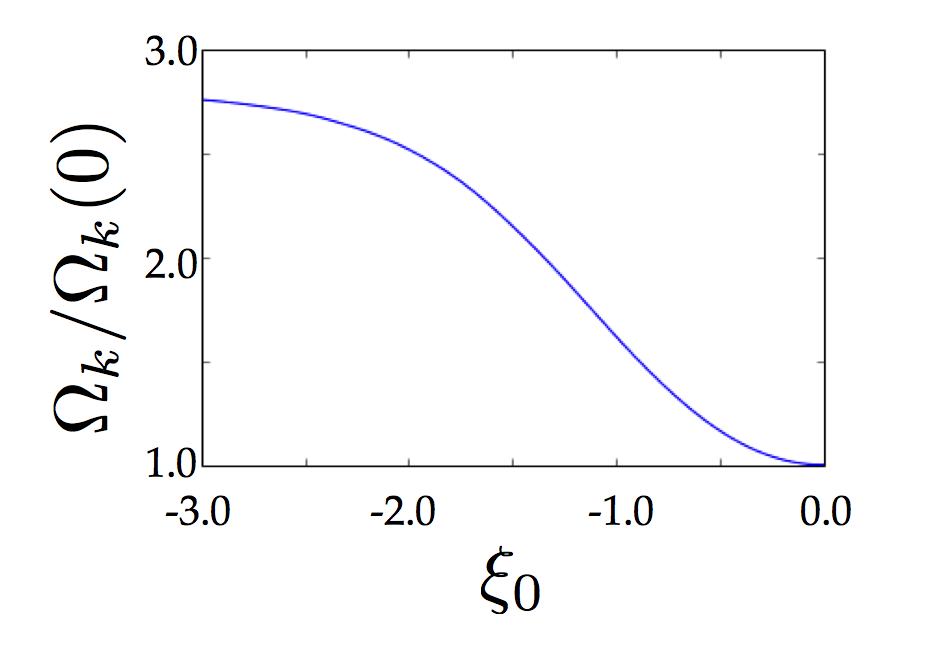}
	\caption{The temperature quadrupole $\mathcal{D}_2$ in $\mu K^2$ (left) and spatial curvature $\Omega_k$ (right) observed at different reference points $\xi_0$ for the AIP model with parameters: $\Delta \phi_1 = 7.1\times 10^{-4}\, \Mpl$, $\beta_1=1.5$, $\sigma=0.004\, \Mpl$ and $\Delta x=1.9 H_F^{-1}$. From the map shown in Eq.~\ref{xix}, $\xi_0=0$ can be identified with the location of the collision. These plots depict only half of the classical transition spacetime. The other half is obtained by mirroring the plots across the right vertical axis.  The anisotropy is largest near the centre of the classical transition, falling off with distance. The spatial curvature appears to asymptote to a constant value a few times larger than its value in the centre of the classical transition, indicating that fewer $e$-folds of inflation occur at the edges of the classical transition than in the centre.}\label{fig:AIPperts}
	\end{center}
\end{figure}

The result for the observed temperature quadrupole and spatial curvature as a function of observer position is shown in Fig.~\ref{fig:AIPperts}. These plots show only half of the classical transition region, which is symmetric about the location of the collision at $\xi_0=0$. The anisotropy is maximized at the location of the collision at $\xi_0 = 0$, falling to zero with increasing $|\xi_0|$. The magnitude of the anisotropy near $\xi_0=0$ is large unless the spatial curvature is relatively small. Comparing to the observed CMB quadrupole, the reference spatial curvature must be of order $\Omega_k(0) \sim 10^{-5}$ in order to produce a temperature quadrupole that is not too large (the observed CMB quadrupole is $\mathcal{D}_2 \simeq 200 \ \mu{\rm K}^2$~\cite{2014A&A...571A..15P}). The spatial curvature increases with $|\xi_0|$, plateauing to a constant value at large $|\xi_0|$.
 
Because of the limited resolution and finite duration of the simulation, it is not possible to extend our description to arbitrarily large $|\xi_0|$. However, the quadrupole drops by roughly three orders of magnitude in the region we can compute. Extrapolating, we can expect that the universe becomes arbitrarily isotropic far from the centre of the classical transition, with fixed spatial curvature. If the curvature perturbation grew with distance from the centre, it would indicate that the surfaces of constant field were going timelike. Since the opposite occurs, there is every indication that the slices of constant field are everywhere spacelike, and therefore, that classical transitions produce  spatially infinite universes. Most of the spatial volume on the constant field slices is far from the location of the collision, where the curvature perturbation vanishes. Therefore, most of the volume to the future of a classical transition is observationally indistinguishable from curved FRW. A similar result was found in Ref.~\cite{wainwright_new}, where it was found that in collisions between non-identical bubbles, infinite surfaces of constant field formed in the region to the future of the collision. The presence of domain walls appears to be the key ingredient in both examples, suggesting that inflating regions within domain walls are generically infinite in spatial extent.

\begin{figure}
\centering
                \includegraphics[width=.32\textwidth]{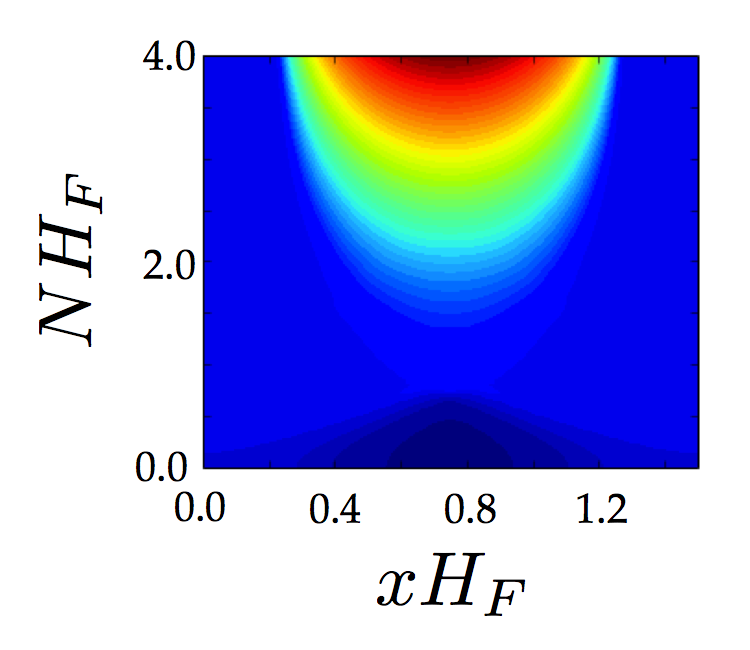}
                \includegraphics[width=.32\textwidth]{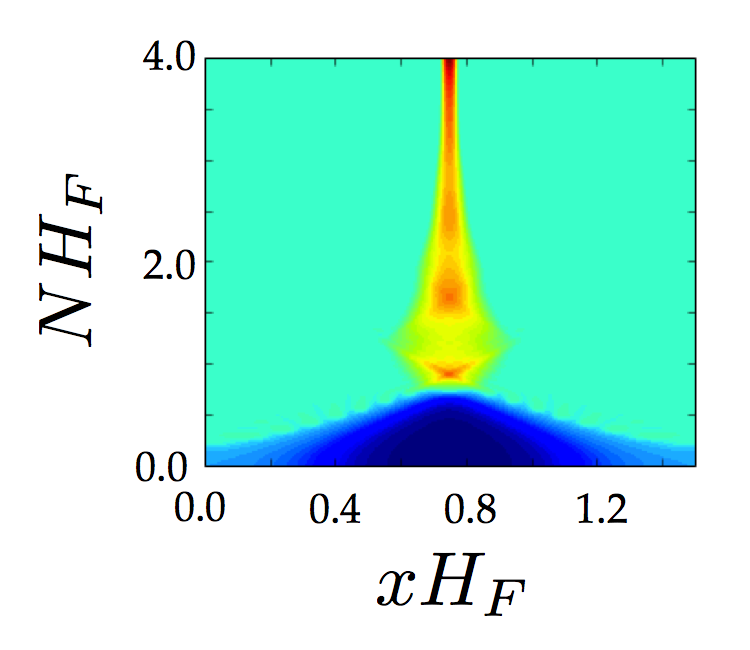}
                \includegraphics[width=.32\textwidth]{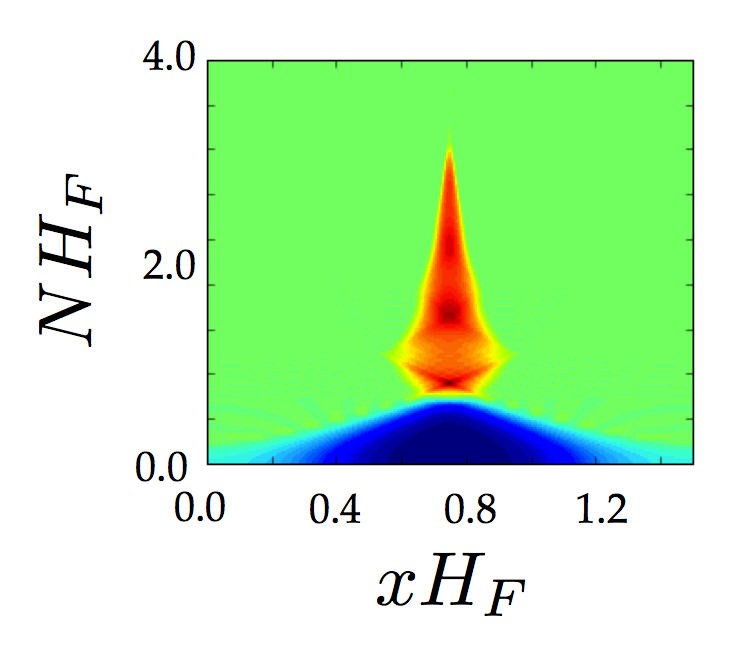}
\caption{Bubble collision simulations for various $\beta_2$. For small $\beta_2$ a Normal geometry is produced (left panel), for intermediate $\beta_2$ a Marginally Repulsive geometry is produced (centre panel), and for a large $\beta_2$ an Oscillatory geometry is produced. See Fig.~\ref{CTfigure} for a detailed summary of the various possible classical transition geometries.}
\label{geobehav}
\end{figure}

Turning to the BIP models, we first confirm that Normal, Oscillatory, and Marginally Repulsive geometries are produced by the simulation. Varying $\beta_2$, which affects the depth of the intermediate vacuum as shown in Fig.~\ref{m6p}, we indeed verify that each geometry can be produced in Fig.~\ref{geobehav}. This is a non-trivial check, as the classical transition produces a universe with a rolling scalar, and not a pure de Sitter space as was assumed in the thin-wall analysis of Ref.~\cite{escapecrunch}.

\begin{figure}
\begin{center}
                \includegraphics[width=.49\textwidth]{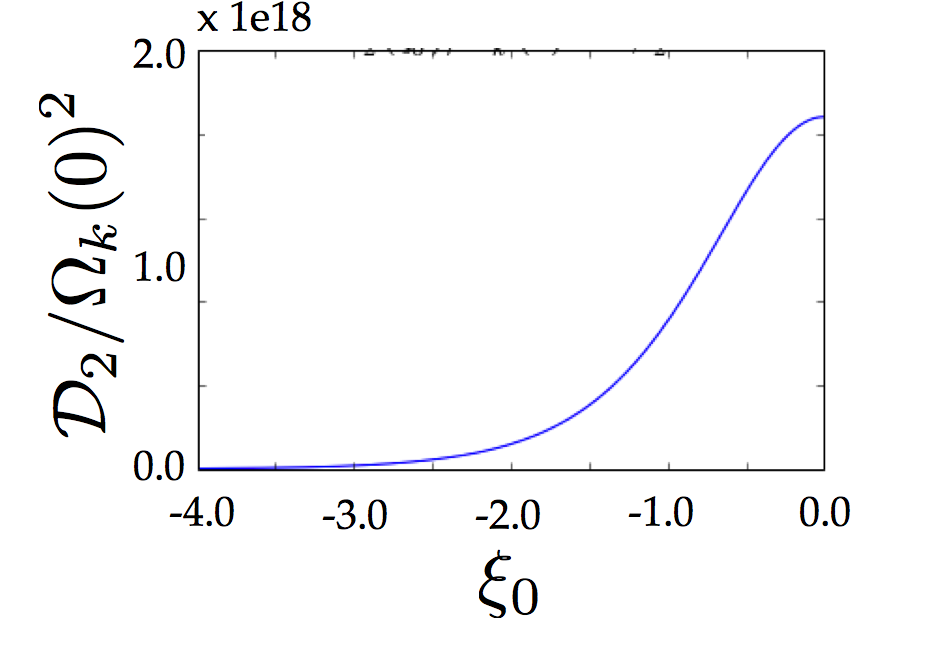}
                \includegraphics[width=.49\textwidth]{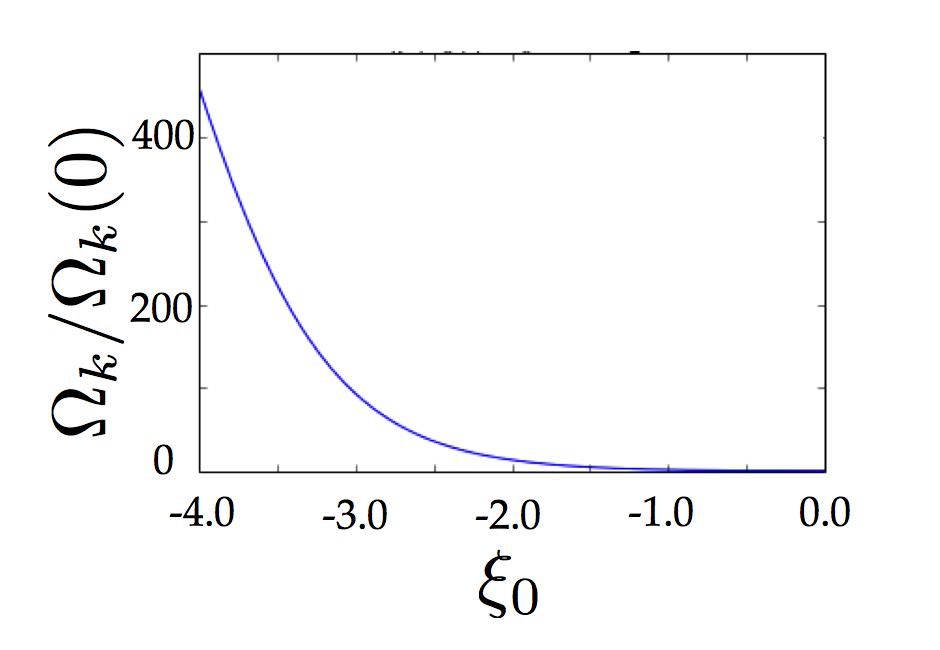}
        \caption{The temperature quadrupole $\mathcal{D}_2$ in $\mu K^2$ (left) and spatial curvature $\Omega_k$ (right) observed at different reference points $\xi_0$ for a marginally repulsive geometry in the BIP model with $\Delta \phi_2 = 3.5 \times 10^{-4} \, \Mpl$, $\beta_2=9.865$ and $\sigma=0.004\, \Mpl$. The anisotropy is largest near the centre of the classical transition, falling off with distance. The spatial curvature increases with $|\xi_0|$, indicating that inflation is significantly disrupted far from the centre of the classical transition.}\label{fig:BIPmodel}
\end{center}
\end{figure}

In the Marginally Repulsive geometry, a lasting inflating region is produced. We use the pipeline described in the previous section to compute observables at different reference points; the result for a model with parameters $\Delta \phi_2 = 3.5 \times 10^{-4} \, \Mpl$, $\beta_2=9.865$ and $\sigma=0.004\, \Mpl$ are shown in Fig.~\ref{fig:BIPmodel}. As for the AIP model presented above, the temperature anisotropies are maximized near the centre of the classical transition region. However, the maximum quadrupole is 6 orders of magnitude larger than the AIP model in this particular example, requiring a far smaller reference curvature to be consistent with the observed quadrupole if we were to inhabit this region. In contrast to the AIP model, the spatial curvature has no signs of plateauing in the region that the simulations can accurately probe. The fall-off of the quadrupole suggests that there is no obstruction to having spatially infinite hyper surfaces. However, the growth of curvature suggests that the Universe near the location of the collision will appear quite different from regions far from the location of the collision. The dramatic growth of curvature could imply negligible inflation far from the location of the collision, although we cannot determine the extent of this effect given the finite resolution of our simulations.

%%%%%%%%%%%%%%%%%%%%%%%%%%%%%%%%%%%%%%%%%%%%%%%
%%%%%%%
%%%%%%%%%%%%%%%%%%%%%%%%%%%%%%%%%%%%%%%%%%%%%%%
%%%%%%%

\subsection{Scanning the parameter space of the AIP model}

As described in Sec.~\ref{secmodel}, for the AIP model we explore slices through parameter space which preserve the radius $R$ and amplitude $\phi_{\rm amp}$ of the bubble walls. In addition, we sample different bubble separations $\Delta x$. Qualitatively, the results over the range of parameters sampled are identical to those described in the previous section. Specifically, the temperature anisotropy peaks at the centre of the classical transition, falling to zero with increasing $|\xi_0|$, and the spatial curvature increases to a constant at large $|\xi_0|$. Variations in the initial separation $\Delta x$ and parameter combinations that keep the amplitude $\phi_{\rm amp}$ fixed but vary the initial radius $R$ control the centre of mass energy in the collision (in flat space, the lorentz factor is given by Eq.~\ref{eq:lorentzfactor}). Parameter combinations that keep the radius $R$ and initial separation $\Delta x$ fixed but vary the amplitude $\phi_{\rm amp}$ are independent of the centre of mass energy, and directly related to the underlying scalar field lagrangian. For a specific potential and initial separation $\Delta x$, the value of $\mathcal{D}_2$ and $\Omega_k$ are correlated at each reference point. We can therefore represent the predictions of a specific model as a line through the $\mathcal{D}_2$-$\Omega_k$ plane. 

\begin{figure}
\begin{center}
                \includegraphics[width=.32\textwidth]{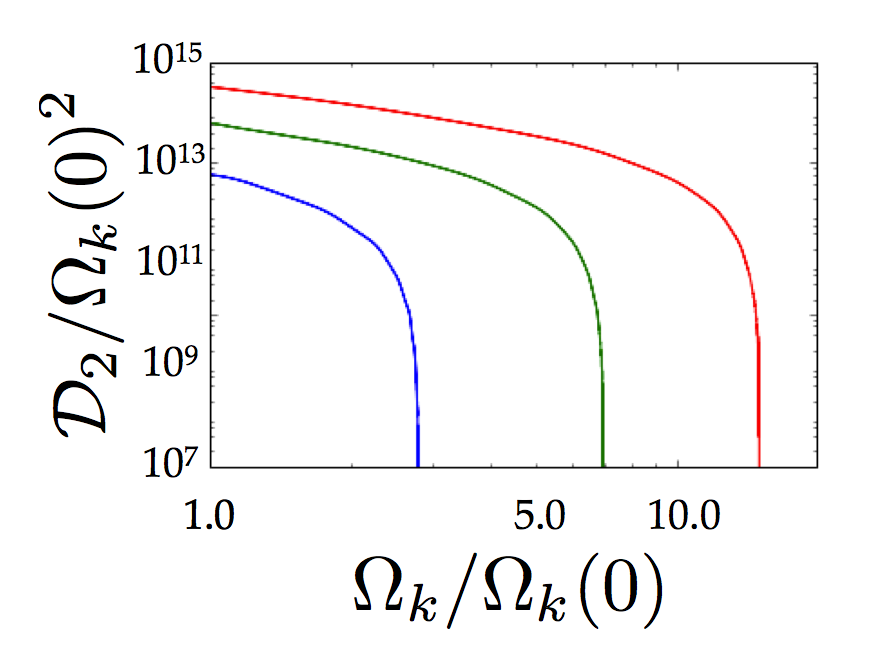}
                \includegraphics[width=.32\textwidth]{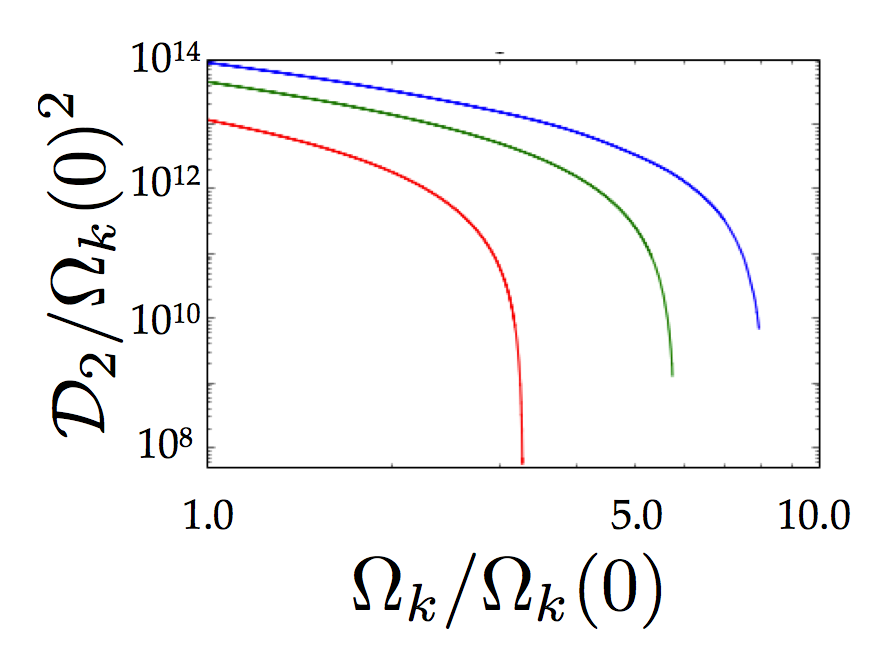}
                 \includegraphics[width=.32\textwidth]{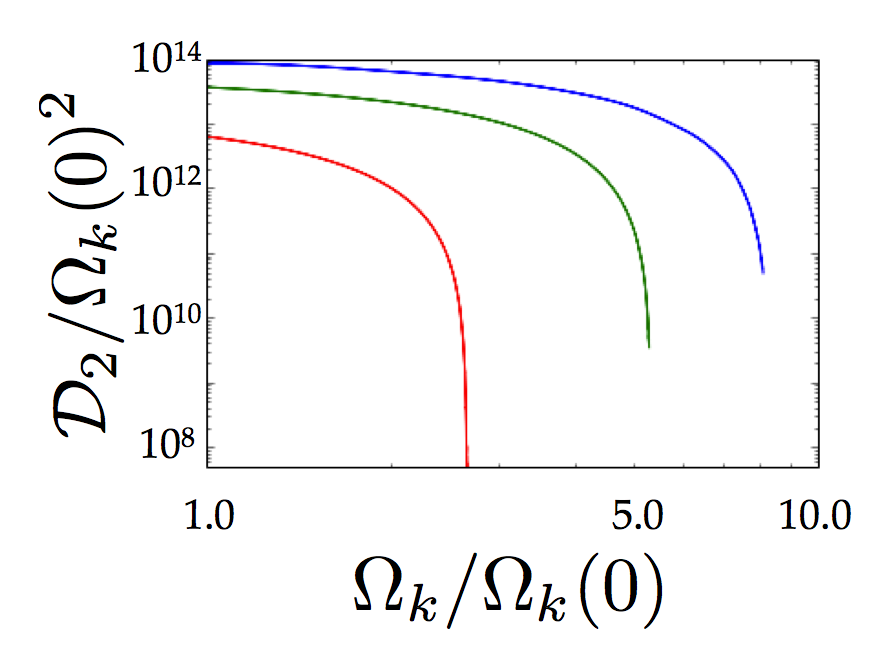}
        \caption{Correlation between $\mathcal{D}_2$ in $\mu K^2$ and $\Omega_k$ in the AIP model. In the left panel, we show variations in the initial separation $\Delta x$. Curves correspond to $\Delta x =1.1$ (blue), and $\Delta x =1.9$ (green), $\Delta x =2.3$ (red). In the centre panel, we show variations in initial radius $R$ from the $\Delta \phi_1$-$\sigma$ sector of the potential. Curves correspond to $H_F R = .246$ (blue), $H_F R = .477$ (green), and $H_F R = .728$ (red). In the right panel, we show variations in initial radius $R$ from the $\beta_1$-$\sigma$ sector of the potential. Curves correspond to $H_F R = .22$ (blue), $H_F R = .53$ (green), and $H_F R = .79$ (red). }\label{eomd2}
\end{center}
\end{figure}

In Fig.~\ref{eomd2} we show the $\mathcal{D}_2$-$\Omega_k$ plane for models where we vary $\Delta x$ and $R$, which control the centre of mass energy of the collision. In the left panel, we see that increasing $\Delta x$ generally leads to larger $\mathcal{D}_2$ and $\Omega_k$. The fall-off of $\mathcal{D}_2$ and plateauing of $\Omega_k$ at large $|\xi_0|$ can be seen as one goes from left to right along each curve. A particular microphysical model (e.g. scalar field potential) yields an ensemble of predictions given by the set of curves in the $\mathcal{D}_2$-$\Omega_k$ plane for all $\Delta x$.  For a specific scalar field lagrangian, bubble collisions will sample all possible values of $\Delta x$ with probability ${\rm Pr} \propto \sin^3 \Delta x$~\cite{statusreport}. In the centre and right panel of Fig.~\ref{eomd2}, we show the predictions in the $\mathcal{D}_2$-$\Omega_k$ plane for varying initial radius $R$, keeping both $\Delta x$ and $\phi_{\rm amp}$ fixed. In general, smaller initial radii (and therefore a higher centre of mass energy) yield larger $\mathcal{D}_2$ and $\Omega_k$. Again, the fall-off of $\mathcal{D}_2$ and plateauing of $\Omega_k$ at large $|\xi_0|$ can be seen as one goes from left to right along each curve. In contrast to the initial separation, the initial radius is a deterministic parameter fixed by the scalar field potential. In conclusion, the observational signatures of a classical transition clearly depend on the centre of mass energy of the collision. More specifically, collisions with a higher centre of mass energy lead to larger predicted values for $\mathcal{D}_2$ and $\Omega_k$.

\begin{figure}
\begin{center}
                \includegraphics[width=.32\textwidth]{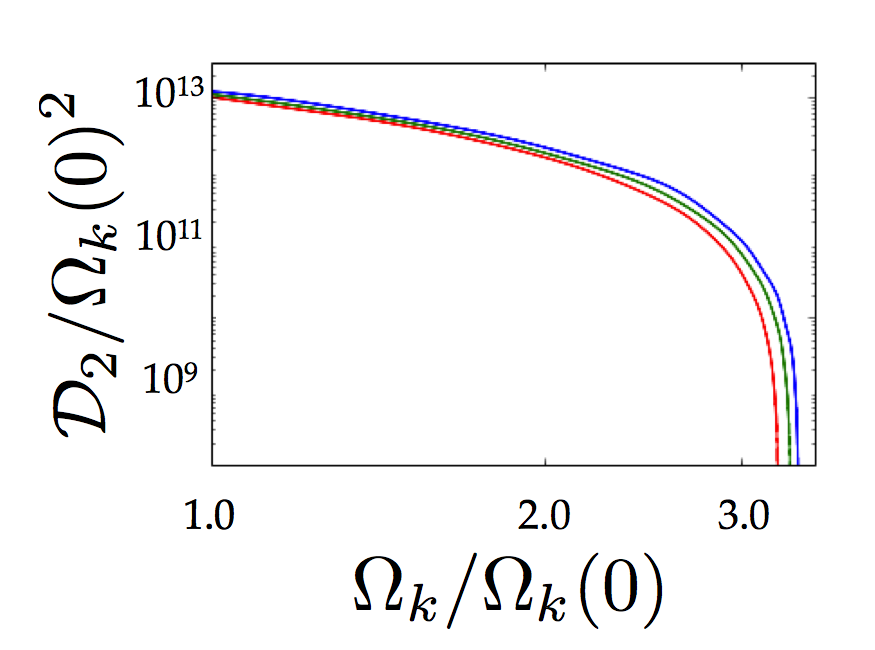}
                \includegraphics[width=.32\textwidth]{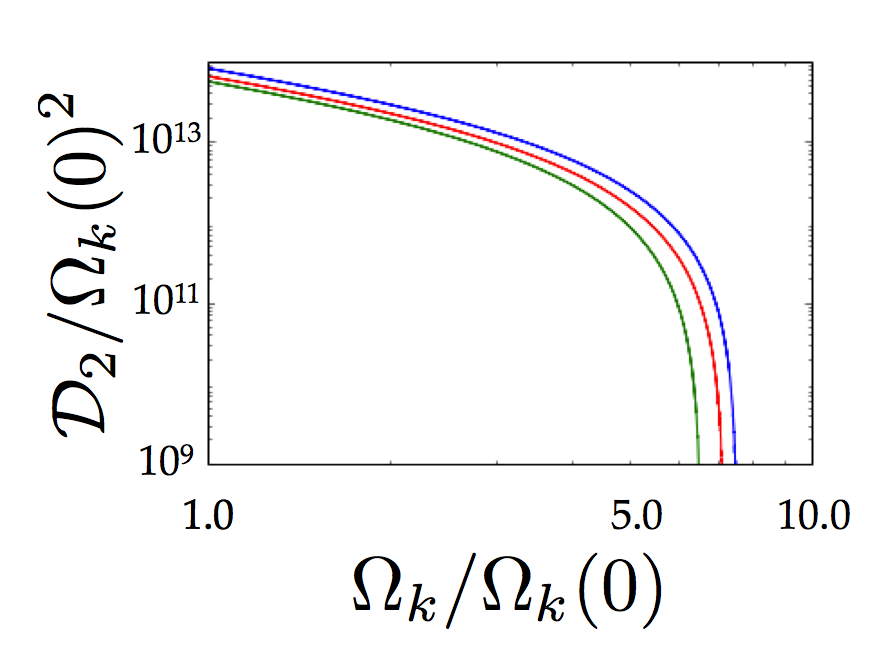}
                \includegraphics[width=.32\textwidth]{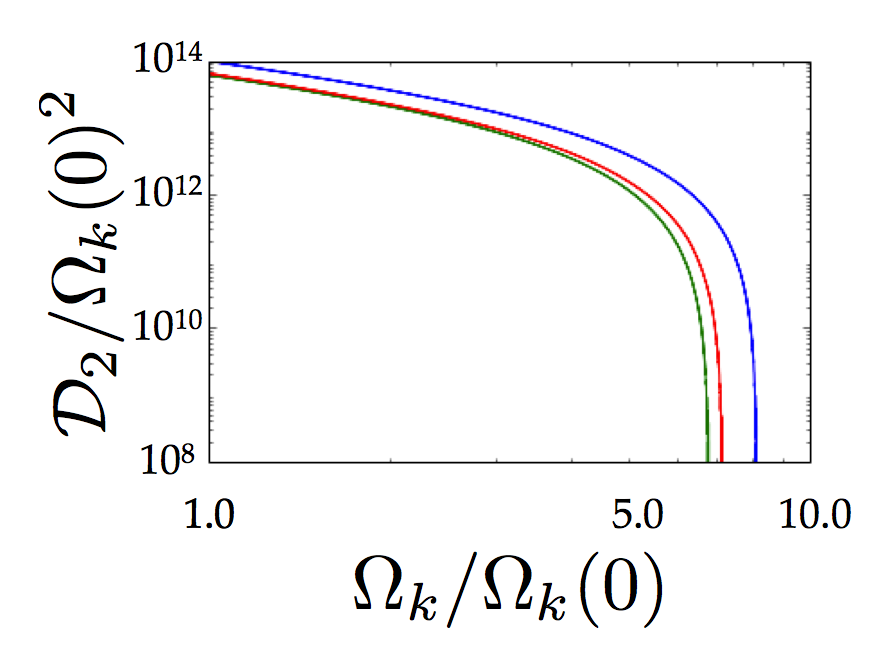}
        \caption{Correlation between $\mathcal{D}_2$ in $\mu K^2$ and $\Omega_k$ in the AIP model. In the left panel, we show variations in the instanton amplitude $\phi_{\rm amp}$ from the $\beta_1$-$\Delta \phi_1$ sector. Curves correspond to $\phi_{\rm amp} = 2.3 \times 10^{-3} \, \Mpl$ (blue), $\phi_{\rm amp} = 2.8 \times 10^{-3} \, \Mpl$ (green), and $\phi_{\rm amp} = 3.3 \times 10^{-3} \, \Mpl$ (red). In the centre panel, we show variations in the instanton amplitude $\phi_{\rm amp}$ from the $\Delta \phi_1$-$\sigma$ sector. Curves correspond to $\phi_{\rm amp} = 1.0 \times 10^{-3} \, \Mpl$ (blue), $\phi_{\rm amp} = 2.2 \times 10^{-3} \, \Mpl$ (green), and $\phi_{\rm amp} = 3.6 \times 10^{-3} \, \Mpl$ (red). In the right panel, we show variations in the instanton amplitude $\phi_{\rm amp}$ from the $\beta_1$-$\sigma$ sector. Curves correspond to $\phi_{\rm amp} = 0.9 \times 10^{-3} \, \Mpl$ (blue), $\phi_{\rm amp} = 1.5 \times 10^{-3} \, \Mpl$ (green), and $\phi_{\rm amp} = 2.5 \times 10^{-3} \, \Mpl$ (red).
}\label{iomd2}
\end{center}
\end{figure}

In Fig.~\ref{iomd2}, we show the $\mathcal{D}_2$-$\Omega_k$ plane for models where the initial separation $\Delta x$ and radius $R$ are  held fixed, but $\phi_{\rm amp}$ varies. Due to computational limitations, we are only able to sample a relatively limited range of $\phi_{\rm amp}$. In the left panel, we vary $\beta_1$ and $\Delta \phi_1$ while holding $\sigma$ fixed at $\sigma=0.006\, \Mpl$. Here, there is an approximately linear decrease in the maximum value of $\mathcal{D}_2$ and an increase in the maximum value of $\Omega_k$. The increase in the maximum value of $\Omega_k$ is consistent with the free-passage approximation, since the field will enter inflation later for a larger $\phi_{\rm amp}$. The decrease in the maximum value of $\mathcal{D}_2$ can be understood as due to the delayed horizon crossing of $\mathcal{R}$, which leads to additional decay. In the $\Delta \phi_1$-$\sigma$ and $\beta_1$-$\sigma$ sectors, there is no apparent trend associated with varying $\phi_{\rm amp}$. This is due to the fact that as $\sigma$ is varied, the field value at which slow-roll begins is varied. 

Returning to the form of the potential (see Fig.~\ref{vhws}), we see that if $\sigma$ is large enough then a period of field evolution will happen before the collision. It was found in Ref.~\cite{Wainwright:2014pta} that field evolution, in addition to the barrier width, can contribute to the field excursion produced by a bubble collision. To explore this behaviour in more detail, we perform simulations in which we vary $\sigma$. For example, choosing $\Delta\phi_{1}=0.0011$, $\Delta\phi_{2}=0.0007$, $\beta_1=2.0$ and $\beta_2=3.0$ a classical transition is obtained for $\sigma < .016$. In this example, the distance the field is displaced by the collision is five times larger than the amplitude of the instanton, which is $\phi_{\rm amp} \simeq .003$. 

\begin{figure}
\centering
                \includegraphics[width=.4 \textwidth]{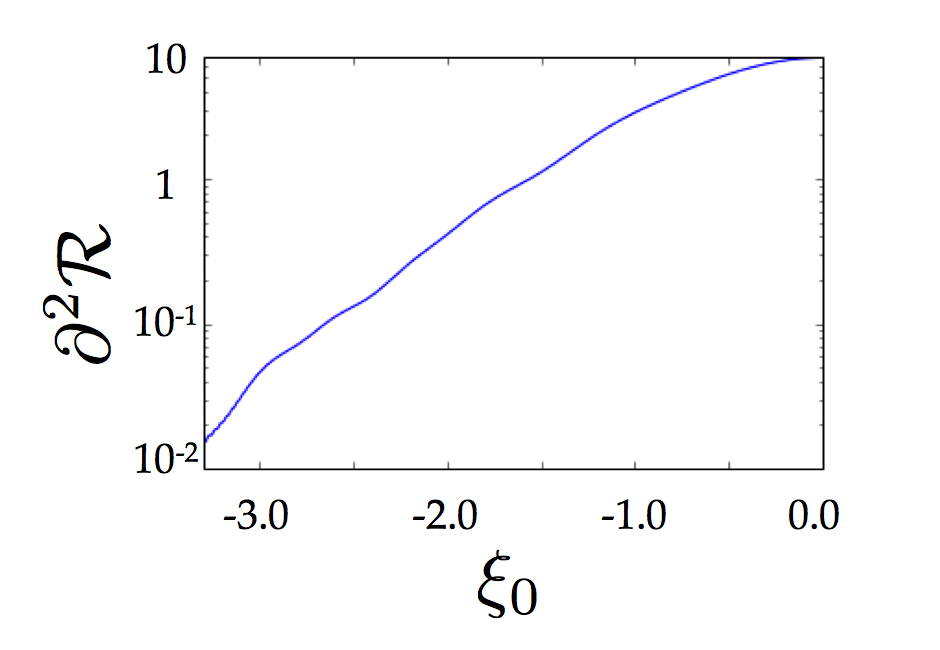}
                \includegraphics[width=.4 \textwidth]{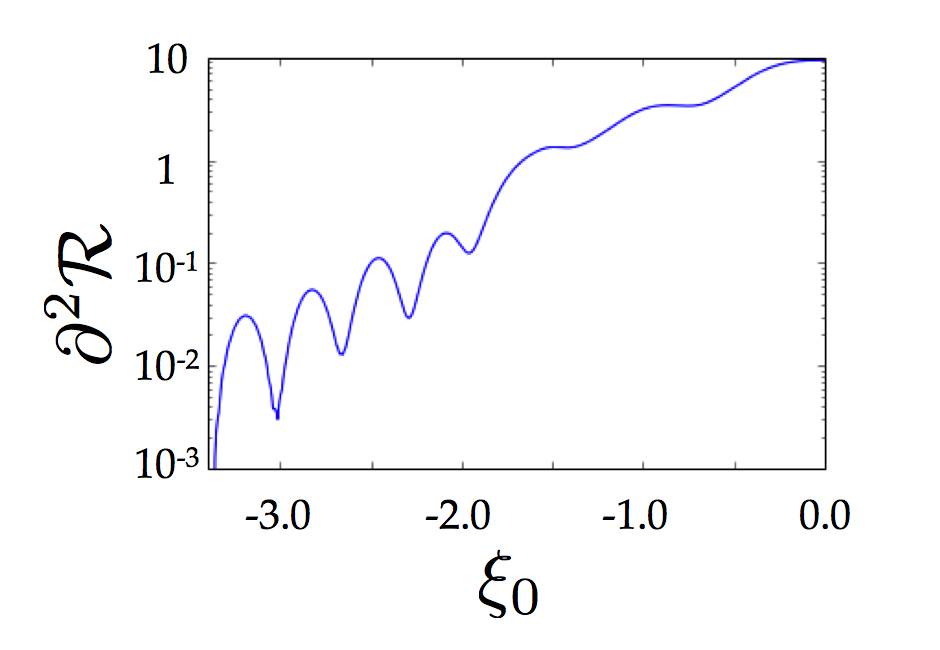}
\caption{Growth of lumpiness as the second bump approaches classical transition boundary. The potential parameters are set to $\Delta\phi_{1}=0.001061\, \Mpl$, $\Delta\phi_{2} = 7.1 \times 10^{-4}\, \Mpl $, $\beta_1=2.0$ and $\beta_2=3.0$. In the left panel, the shift is set to $\sigma=.007$, where there are few wall modes excited. In the right panel, the shift is set to $\sigma=.016$, just at the edge of the range where classical transitions are possible. The wall modes are visible as oscillations in the comoving curvature perturbation.}\label{nctb}
\end{figure}

Adjusting $\sigma$ such that we approach the boundary where classical transitions are allowed, the comoving curvature perturbation develops additional structure. In Fig.~\ref{nctb} we show $\partial^2_\xi \mathcal{R}$, which develops an oscillatory structure as the classical transition boundary is approached. These features are produced by the excitation of the internal degrees of freedom of the post-collision domain walls. Such wall modes were also observed in Ref.~\cite{wainwright_new} for the collision between non-identical bubbles. Wall modes add extra structure to the comoving curvature perturbation beyond the quadrupole. We leave an exploration of the phenomenological consequences of wall modes to future work.

%%%%%%%%%%%%%%%%%%%%%%%%%%%%%%%%%%%%%%%%
%%%%%%
%%%%%%
%%%%%%%%%%%%%%%%%%%%%%%%%%%%%%%%%%%%%%%%
%%%%%%
%%%%%%
%%%%%%%%%%%%%%%%%%%%%%%%%%%%%%%%%%%%%%%%
%%%%%%
%%%%%%

\subsection{Exploring the BIP model}

\begin{figure}
\centering
\includegraphics[width=.65\textwidth]{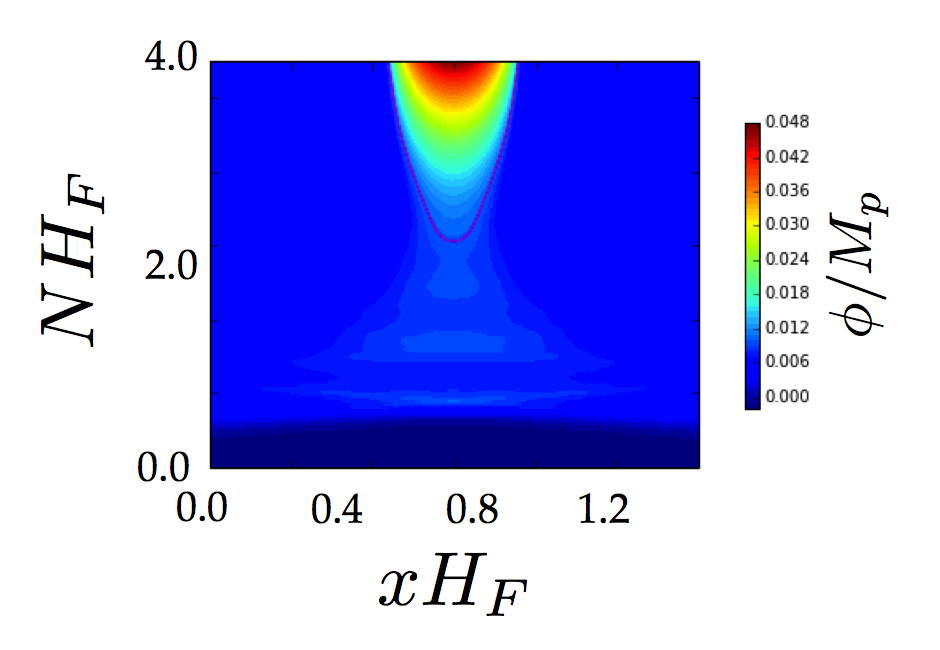}
\caption{An example of countour plot for the second model in the $\beta_2-\sigma$ sector. The magenta line indicates the first space like constant-field hyper surfaces, which we take as the beginning of inflation.}\label{m6ctsb}
\end{figure}

A marginally repulsive geometry is produced for a specific relation between the vacuum energies and the energy density at the onset of slow-roll inflation, as described in Sec.~\ref{sec:classtrans}. The vacuum energies of the two minima are unambiguous. However, we must impose some prescription for measuring the energy density at the onset of slow-roll. We define the onset of slow-roll as the field value $\phi_\mathrm{sp}$ along the first space like surface of constant field; an example is shown in Fig.~\ref{m6ctsb}. 

\begin{figure}
\begin{center}
                \includegraphics[width=.48\textwidth]{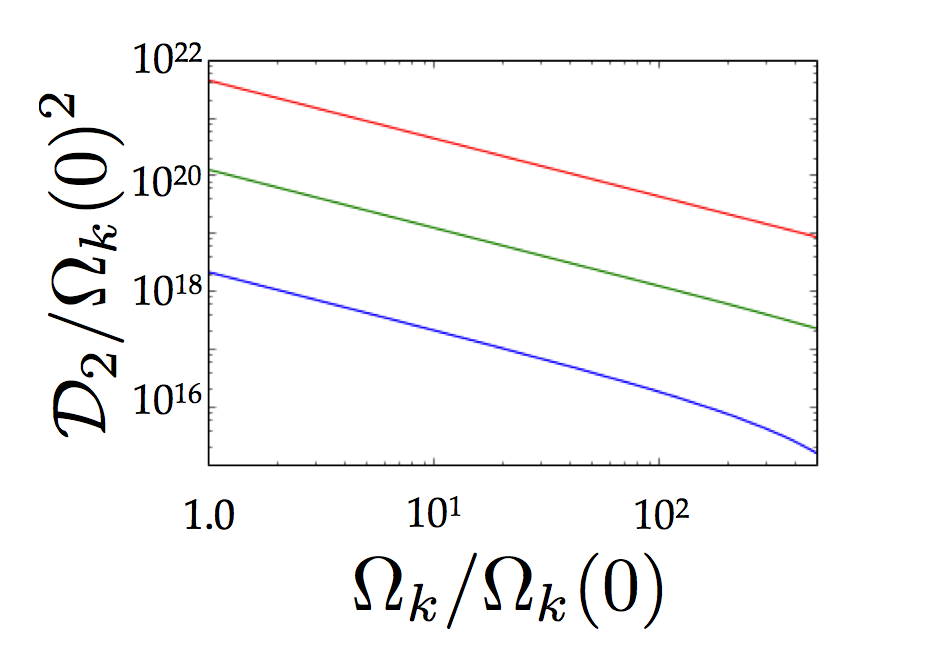}
                \includegraphics[width=.48\textwidth]{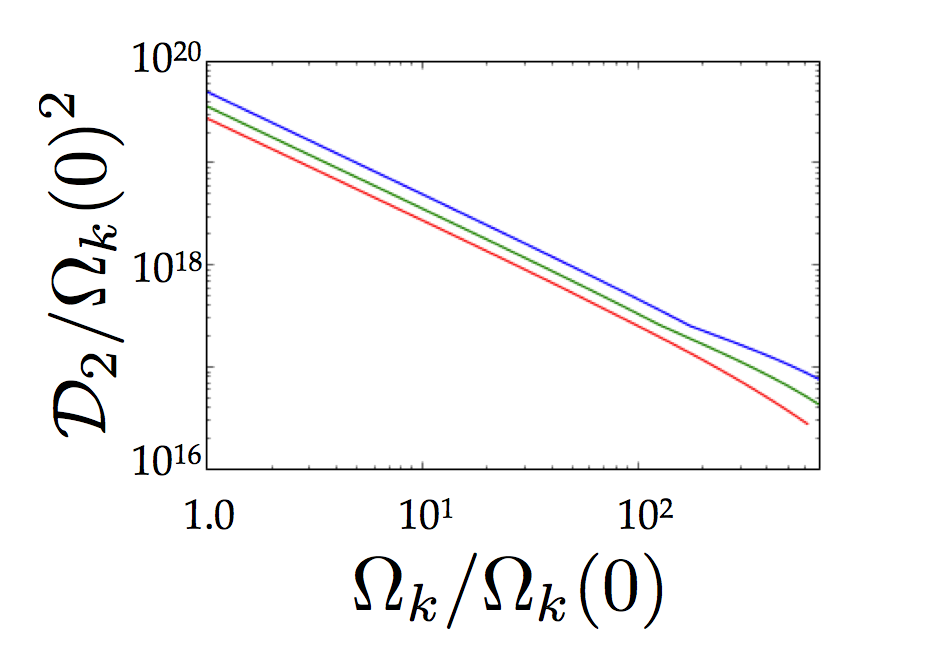}
        \caption{Correlation between $\mathcal{D}_2$ in $\mu K^2$ and $\Omega_k$ for the BIP model. In the left panel we sample different values of $\phi_{\rm sp}$ by varying parameters in the $\sigma-\beta_2$ sector. Curves correspond to $\phi_{\rm sp} = 0.01\, \Mpl$ (blue), $\phi_{\rm sp} = 0.018\, \Mpl$ (green), and $\phi_{\rm sp} = 0.025\, \Mpl$ (red). In the right panel we sample different values of $\phi_{\rm sp}$ by varying parameters in the $\Delta\phi_2-\beta_2$ sector. Curves correspond to $\phi_{\rm sp} = 0.0130\, \Mpl$ (blue), $\phi_{\rm sp} = 0.0121\, \Mpl$ (green), and $\phi_{\rm sp} = 0.0114\, \Mpl$ (red).}\label{omd2m6}
        \end{center}
\end{figure}

In Fig.~\ref{omd2m6} we show the $\mathcal{D}_2$-$\Omega_k$ plane in the two sectors ($\sigma-\beta_2$ and $\Delta\phi_2-\beta_2$) of parameter space that give marginally repulsive classical transition geometries. Comparing with the results from the AIP model shown in Figs.~\ref{eomd2} and \ref{iomd2}, there are a number of notable differences. First, the size of both $\mathcal{D}_2$ and $\Omega_k$ is orders of magnitude larger in the BIP models. In addition, there is no clear asymptote of $\Omega_k$ at large $\xi_0$. Looking at the trends associated with the field value $\phi_\mathrm{sp}$ at the onset of slow-roll, we see that the signature grows with increasing $\phi_\mathrm{sp}$. This makes intuitive sense, as more violent initial perturbations will delay the onset of inflation. Finally, we have explicitly verified that the energy density at the onset of inflation is above the energy density at the intermediate potential medium, but obeys the bound Eq.~\ref{eq:mrceq}.

%%%%%%%%%%%%%%%%%%%%%%%%%%%%%%%%%%%%%%%%%%%%%%
%%%%%%
%%%%%%%%%%%%%%%%%%%%%%%%%%%%%%%%%%%%%%%%%%%%%%
%%%%%%

\section{Conclusion}\label{secconclude}

In this work, we investigated the possibility of observationally determining whether our observable Universe arose from a classical transition produced by the collision of two bubbles. We consider single-field models with a potential that has three local minima, as depicted in Fig.~\ref{CTfigure}. To have a viable model of our observable Universe, we additionally require a slow-roll plateau leading to the lowest energy minimum. There are two classes of models that could give rise to our observable Universe. The first class of models (AIP models) produce Normal classical transition geometries. These models have three minima with sequentially decreasing energy density, and the collision produces an ever-expanding region containing the lowest energy state. The second class of models (BIP models) produce Marginally Repulsive classical transition geometries. In the BIP models the classical transition produces an inflationary epoch of a higher energy than that of the bubble interiors (but obeying the bound Eq.~\ref{eq:mrceq}). 

In both models, we have found that classical transitions produce approximately FRW Universes with negative spatial curvature that are (extrapolating our simulation) infinite in extent. The classical transition geometry is inhomogeneous, but retains approximate planar symmetry. We expand the classical transition geometry into local approximately FRW patches characterized by the local spatial curvature and the comoving curvature perturbation. The leading observables in such a Universe are a CMB quadrupole and negative spatial curvature. 

Our main results are presented in Figs.~\ref{eomd2},~\ref{iomd2}, and~\ref{omd2m6} which show the the $\mathcal{D}_2$-$\Omega_k$ plane for varying parameters in the models under consideration. All of our results are presented as a function of an arbitrary reference curvature $\Omega_k (0)$, which depends on the details of slow-roll and reheating. In the AIP model, the spatial curvature increases to a constant with distance from the centre of the classical transition. The locally observed quadrupole drops to zero with distance from the centre. The locally observed quadrupole also drops to zero with distance from the centre of the classical transition in the BIP model, however, the spatial curvature appears to increase without bound.  

The magnitude of the locally observed quadrupole and negative spatial curvature depends both on the centre of mass energy of the collision (fixed by the initial separation $\Delta x$ and initial radius $R$ of the colliding bubbles) and the scalar field lagrangian. In both the AIP and BIP models, decreasing initial radius or increasing initial separation generally leads to a larger locally observed quadrupole and curvature. Varying parameters in the scalar field lagrangian, we choose parameter combinations in the AIP model that fix the initial radius of the bubble, but change the total field excursion $\phi_{\rm amp}$ between the instanton endpoints. We find that an increasing amplitude leads to a larger locally observed quadrupole and curvature. We have also found that motion of the field after tunnelling can contribute significantly to the distance in field space that the classical transition can reach. Finally, we have found that the amplitude of the locally observed quadrupole and curvature is generally far larger in the BIP models than the AIP models.

Although we have not investigated how changes in the inflationary potential will affect our results, in analogy with previous results on bubble collisions~\cite{Wainwright:2014pta}, we can expect that small-field models of inflation to be more susceptible to disruption than large-field models of inflation~\footnote{Small field models are defined as those in which inflation takes place over a sub-planckian field range, while large field models are defined as those in which inflation takes place over a super-planckian field range.}. In the moments after a collision, the details of the potential are irrelevant. Therefore, we can compare the field excursion predicted by the free passage approximation to the size of the region in field space over which inflation occurs. For a fixed barrier structure, the field excursion produced by the classical transition is fixed, and so small-field models will experience a larger fractional displacement than large-field models. The field excursion during inflation is related to the tensor to scalar ratio $r$ by the Lyth bound~\cite{Lyth:1996im}: $\Delta \phi / \Mpl \simeq \sqrt{r/0.01}$. For a fixed barrier structure, we therefore expect that the duration of inflation will be affected more severely in models with small $r$. The duration of inflation in a given model (modulo the details of reheating) determines the reference curvature $\Omega_k(0)$, which in turn determines the magnitude of observables. Therefore, inflationary models with small $r$ can be expected to yield a larger spatial curvature and a larger locally observed quadrupole. 

Will we ever be able to test whether or not our observable Universe was created by a classical transition? The fact that the observables are only curvature and a quadrupole makes it difficult to confirm whether or not a classical transition in a single field model occurred in our past. Curvature can be produced in many scenarios, and the temperature quadrupole has an enormous error bar due to cosmic variance. Ruling out models with a classical transition requires an observation of spatial curvature. Current observation constrains curvature at the level of $\Omega_k \simeq 0.000 \pm .005 $. The confusion limit with super horizon perturbations is $\Omega_k \sim \pm 10^{-5}$, which represents a lower bound on what level of primordial curvature could ever be observed. An observation of positive spatial curvature at a level greater than this would rule out a classical transition in our past. If an observation of negative spatial curvature is made, one can look at the correlated quadrupole in a given model, and compare to the observed value. However, in the AIP models, a particular value of the spatial curvature can be correlated with arbitrarily small values of $\mathcal{D}_2$ due to the fact that curvature asymptotes to a constant at large distance from the centre of the classical transition, while the locally observed quadrupole drops to zero. In the BIP models, the curvature grows with distance from the centre while the quadrupole shrinks. Therefore, for a fixed $\Omega_k(0)$ it is conceivable that one could rule out some portion of model space based on the non-observation of a large curvature or a large quadrupole, since one or the other will be predicted depending on the location of the observer. However, $\Omega_k(0)$ is dependent on the number of $e$-folds in the region that underwent a classical transition, which is completely unconstrained. This makes it impossible to conclusively rule out a classical transition in our past. 

Signatures in multi-field models could be more varied than single field models, for example giving rise to isocurvature perturbations or leading to new sources of instability (see e.g.~\cite{Braden:2014cra,Braden:2015vza,Bond:2015zfa}). In addition, a deeper exploration of the role of wall modes is warranted, and may yield a more distinctive signature than a CMB quadrupole. This may make it easier to constrain models where our Universe arose from a classical transition. In future work, we plan to explore the phenomenology of bubble collisions and classical transitions in multi field models. It would also be desirable to embed a model of classical transitions in a more realistic theory, for example string theory. Indeed, classical transitions may be expected in the context of flux compactifications~\cite{Deskins:2012tj}. Regardless, this paper illustrates that models of eternal inflation can make predictions for real observables, adding further credence to a scientific theory of the multiverse.

\acknowledgments

Research at Perimeter Institute is supported by the Government of Canada through Industry Canada and by the Province of Ontario through the Ministry of Research and Innovation. MCJ is supported by the National Science and Engineering Research Council through a Discovery grant. This work was supported in part by National Science Foundation Grant No. PHYS-1066293 and the hospitality of the Aspen Center for Physics.

%%%%%%%%%%%%%%%%%%%%%%%%%%%%%%%%%%%%%%%%%%%%%%
%%%%%%
%\label{Bibliography}
%%\bibliographystyle{plainnat}
%\bibliography{bcct}

\bibliographystyle{JHEP}
\bibliography{bcct}

\end{document}